\documentclass[aps,prd,twocolumn,showpacs,nofootinbib,eqsecnum,superscriptaddress,longbibliography]{revtex4-1}

\usepackage[centertags]{amsmath}
\usepackage{amssymb}
\usepackage{latexsym}
\usepackage{enumerate}
\usepackage{graphicx}
\usepackage{mathrsfs}
\usepackage{hyperref}
\usepackage{stmaryrd}
\usepackage{import}
\usepackage{tensor}
\usepackage{xcolor}
\definecolor{colour1}{HTML}{0571b0}
\hypersetup{colorlinks=true, linkcolor=colour1, citecolor=colour1,
  filecolor=colour1, urlcolor=colour1, pdfencoding=auto}
\makeatletter
\def\@bibdataout@aps{\immediate\write\@bibdataout{%
@CONTROL{apsrev41Control,author="08",editor="1",pages="1",title="0",year="0"}}%
  \if@filesw\immediate\write\@auxout{\string\citation{apsrev41Control}}\fi}
\makeatother
\def\doibase{https://doi.org/}
\usepackage{bm}
\usepackage{multirow}
\usepackage{breakurl}
\usepackage{float}
\usepackage{lipsum}
\usepackage{siunitx}
\usepackage{comment}
\usepackage{mathtools}
\usepackage{autobreak}
\usepackage[caption=false]{subfig}

\mathtoolsset{centercolon}

\usepackage{stmaryrd}
\usepackage{color}
\usepackage{amsmath}
\usepackage{amssymb}
\usepackage{latexsym}
\usepackage{enumerate}
\usepackage{graphicx}
\usepackage{mathrsfs}
\usepackage{physics}
\usepackage{tabularx}
\usepackage{mathtools}
\makeindex

\newcommand{\be}{\begin{equation}}
\newcommand{\ee}{\end{equation}}
\newcommand{\bea}{\begin{eqnarray}}
\newcommand{\eea}{\end{eqnarray}}
\newcommand{\ba}{\begin{array}}
\newcommand{\ea}{\end{array}}
\newcommand{\bi}{\begin{itemize}}
\newcommand{\ei}{\end{itemize}}

\renewcommand{\l}{\left}
\renewcommand{\r}{\right}
\renewcommand{\a}{\alpha}
\renewcommand{\b}{\beta}
\newcommand{\g}{\gamma}
\newcommand{\G}{\Gamma}
\renewcommand{\d}{\delta}
\newcommand{\D}{\Delta}

\newcommand{\eps}{\epsilon}

\newcommand{\ka}{\kappa}
\newcommand{\la}{\lambda}
\newcommand{\vp}{\varphi}
\renewcommand{\O}{\Omega}
\renewcommand{\o}{\omega}
\renewcommand{\th}{\theta}

\newcommand{\q}{\quad}
\renewcommand{\qq}{\qquad}

\newcommand{\Sig}{\Sigma}

\newcommand{\oo}{\infty}

\allowdisplaybreaks

\begin{document}
    
\title{Post-Newtonian expansion of the energy and angular momentum flux radiated to infinity: Eccentric-equatorial 
orbits about a Kerr black hole}

\author{Jezreel C. Castillo}
\affiliation{School of Mathematical Sciences and STAG Research Centre, University of Southampton, 
Southampton SO17 1BJ, United Kingdom}
\affiliation{Department of Physics and Astronomy, University of North Carolina at Chapel Hill, Chapel Hill, 
North Carolina 27599} 
\author{Christopher Munna}
\affiliation{Department of Physics and Astronomy, University of North Carolina at Chapel Hill, Chapel Hill, 
North Carolina 27599} 
\author{Zachary Nasipak}
\affiliation{School of Mathematical Sciences and STAG Research Centre, University of Southampton, 
Southampton SO17 1BJ, United Kingdom}
\affiliation{Department of Physics and Astronomy, University of North Carolina at Chapel Hill, Chapel Hill, 
North Carolina 27599}
\author{J. David Brown}
\affiliation{Department of Physics, North Carolina State University, Raleigh, North Carolina 27695}
\author{Charles R. Evans}
\affiliation{Department of Physics and Astronomy, University of North Carolina at Chapel Hill, Chapel Hill, 
North Carolina 27599}
\affiliation{School of Mathematics and Statistics, University College Dublin, Belfield D04 N2E5, Dublin 4, Ireland}

\begin{abstract}

We present analytical expressions for the energy and angular momentum fluxes radiated to infinity by a (nonspinning) 
point mass in an eccentric-equatorial orbit about a Kerr black hole.  These results build upon our previous work 
involving point masses on eccentric orbits about a Schwarzschild black hole.  We now adopt the Teukolsky formalism 
in first-order black hole perturbation theory to extend our methods to Kerr.  We obtain the expressions using the 
method of Mano, Suzuki, and Takasugi, which provides infinite-series analytical solutions to the Teukolsky equation 
that we generate using \textsc{Mathematica}.  The expressions presented are obtained as a double expansion in the 
post-Newtonian (PN) compactness parameter $1/p$, where $p$ is the dimensionless semi-latus rectum, and in the 
eccentricity $e$, up to 10PN and $e^{20}$, extending beyond prior 6PN and $e^{16}$ results available for generic 
orbits.  We characterize how the PN flux coefficients depend on the black hole spin $a$ and $e$, and find a set of
spin-dependent components that can be resummed as closed-form expressions in $e$.  The analytical expressions are 
then validated with numerical comparisons against high-precision sample results from an independent Teukolsky code 
and with a broader set of numerically interpolated values derived from a flux data set produced with the same 
code.  For $a=0.9M$, the numerically evaluated 10PN fluxes match current interpolated flux grids ($\sim10^{-7}$) 
for $p\gtrsim20$--$75$ as $e$ ranges from $0.1$ to $0.7$, where they could replace Teukolsky-code results.

\end{abstract}

\maketitle

\section{Introduction}

Millihertz gravitational wave sources, such as massive black hole binaries, galactic compact binaries, and
extreme-mass-ratio inspirals \cite{AmarETC13}, will soon be observable by space-based detectors.  Leading this effort 
is the Laser Interferometer Space Antenna (LISA) \cite{ColpETC24,LISA}, planned for launch in 2035, with the proposed 
Taiji \cite{LuoETC21} and TianQin \cite{LuoETC16} missions expected on a similar timescale. Of particular interest are
extreme-mass-ratio inspirals (EMRIs), binaries in which a stellar-mass compact object of mass $\mu \sim 10 M_\odot$
inspirals into a massive black hole of mass $M \sim 10^6 M_\odot$, so that $\mu/M \ll 1$
\cite{AmarETC07,BaraETC18,BerrETC19}. EMRIs will spend months to years in the LISA band, and their signals carry rich
structure imprinted by the multi-periodic dynamics of the orbit. These features will enable high-precision measurements
of EMRI source parameters \cite{BabaETC17,SperETC26}.

The detection and characterization of gravitational signals require sufficiently accurate waveform models 
that are built from solutions to Einstein's equations.  Two widely used schemes for solving Einstein's 
equations are black hole perturbation theory (BHPT) and post-Newtonian (PN) theory.  BHPT builds solutions by 
expanding in the mass ratio $\mu/M$ and is applicable when the component bodies have disparate masses 
\cite{BaraPoun18,PounWard22}, making it a natural framework for modeling EMRIs. PN theory expands in the 
(dimensionless) velocity $v/c$ and is applicable when the component bodies are widely separated and slowly moving 
\cite{Blan14}.

For EMRIs these two regimes of applicability are not mutually exclusive but coincide during the early inspiral 
phase.  This leads to a region of overlap where $\mu/M \ll 1$ and $v/c \ll 1$ 
\cite{Detw08,SagoBaraDetw08,BaraSago09,BlanETC09,BlanETC10}, and researchers often exploit this shared domain to 
compare independent results from BHPT and PN. The overlap regime also allows for the results of one formalism to inform
the other. In particular, there has been steady progress over recent decades in extracting high-order PN
expansions of self-force quantities---such as the time-averaged gravitational wave fluxes
\cite{GanzETC07,SagoFuji15,FujiShib20,IsoyETC22,ForsEvanHopp16,MunnETC20,Munn20,MunnEvanFors23,CastETC25,SagoETC26},
generalized redshift invariant \cite{BiniDamoGera15,BiniDamoGera16b,BiniGera19a,MunnEvan22a,Munn23}, spin-precession
invariant \cite{BiniGera19b,MunnEvan22b} and even the gravitational waveforms \cite{IsoyETC22}---at first order in the
mass ratio. The analytic expansions of these quantities serve as practical inputs for effective-one-body
\cite{DamoNaga07, Damo10, YuneETC11, TaraETC14b, NagaAlba22, AlbaETC23, LeatBuonVand25} and hybrid waveform models
\cite{HuerETC18,BurkETC24,PaulETC25,HonePounComp26}, while in self-force-based models \cite{ChapBirdETC25, MathETC26}
these high-order expansions can serve in place of numerical data where such data are not yet available or are less
accurate (i.e., at large separations).

Much of this progress has been made possible by the work of Mano, Suzuki, and Takasugi (MST) 
\cite{ManoSuzuTaka96a, ManoSuzuTaka96b}, in which they developed a method for solving the homogeneous first-order 
perturbation equations via infinite convergent sums of analytic functions \cite{ManoSuzuTaka96a,SasaTago03}.  PN 
expansions are extracted in two ways, either through a numerical fitting scheme involving highly accurate numerical 
output \cite{ShahFrieWhit14,Shah14,ForsEvanHopp16, MunnETC20} or through a direct symbolic expansion using select 
PN parameters \cite{BiniDamo13,BiniDamo14a,BiniDamo14b, BiniDamo14c,KavaOtteWard15,KavaOtteWard16,Munn20,CastETC25,
CastETC25b,NeefKavaOtte26}.  The full inhomogeneous solution can then be constructed by joining the homogeneous 
solutions to a PN expansion of the source terms.

In this work we use this MST method to calculate the PN expansions of the time-averaged energy and angular momentum 
flux radiated to infinity by a (non-spinning) point particle on an eccentric-equatorial geodesic about a rotating 
massive black hole.  This builds on previous efforts of
\cite{ForsEvanHopp16,MunnEvan19a,MunnETC20,MunnEvan20a,Munn20,Munn20c}, in which the MST solutions were used within the
Regge-Wheeler-Zerilli (RWZ) formalism to extract detailed knowledge of the fluxes for eccentric Schwarzschild EMRIs.
In that work, each PN term was also expanded in the eccentricity $e$ of the orbit.  However, developments in the 
full PN theory show that some of these (non-hereditary) PN terms should correspond to closed-form or compact 
functions of $e$.  Accordingly, an important part of the computation is an attempt to manipulate or resum the 
eccentricity Taylor series into underlying compact expressions where possible.  Early work 
\cite{ForsEvanHopp16, MunnETC20} calculated fluxes numerically to very high precision (hundreds of digits) and 
then fitted the expected form of the PN expansion.  The resulting highly accurate numerical values of the 
coefficients in the eccentricity expansion could then be converted to analytic form in many cases using an integer 
relation algorithm (e.g., PSLQ \cite{FergBailArno99}).  Later work \cite{Munn20} moved to computing these expansions 
fully symbolically for fluxes at infinity through 19PN and $e^{10}$, and through 10PN and $e^{20}$.  A summary of 
all these results is provided in Table~IV of \cite{Munn20}.  As these results were being developed and once enough 
terms in the eccentricity expansion of some enhancement functions were revealed, it was possible to see patterns 
that led to natural resummations.  In \cite{MunnEvan19a,MunnEvan20a}, these expansions were combined with exact 
results on low-order multipole moments from full PN theory to derive infinite series with exact coefficients for 
an infinite sequence of flux terms known as leading logarithms and their 1PN corrections.  The leading logarithm 
series refers to the first integer- and half-integer-order appearances of each new power of the log of the 
compactness parameter in the PN series.  The $k$PN logarithm series corresponds to the $k$th PN correction to the 
leading logs.  

Due to the added difficulty of incorporating the spin parameter $a$, analytical calculations involving generic orbits 
in Kerr have historically seen less development \cite{SagoFuji15,FujiShib20,IsoyETC22}, with the state of the art
reaching 6PN and $e^{16}$ \cite{SagoETC26}.  As with previous works, we find the addition of $a$ significantly 
increases the complexity of the expansion.  

Nevertheless, we are able to apply the direct analytical expansion method in \cite{Munn20}, modified for use with
the Teukolsky MST solutions \cite{CastETC25,CastETC25b}, as was previously done for the generalized redshift
invariant on eccentric-equatorial Kerr orbits \cite{Munn23}.  Key recent simplifications are to calculate Coulomb wave
functions, represent the near-horizon downgoing-wave solution as a linear combination of them, and scale the 
mode functions by an $r$-independent amplitude, all of which speed up the symbolic calculations.  These allow 
us to derive the energy and angular momentum fluxes radiated toward infinity to 10PN and $e^{20}$, which we present 
later in this paper.  The calculation was carried out independently by two of us (JCC and CM), in separate
\textsc{Mathematica} codes, one based on the $s=+2$ Teukolsky function (related to $\psi_0$) and the other on
$s=-2$ (related to $\psi_4$).  The two sets of flux expansions were found to agree exactly through 8PN and $e^{20}$,
which was the original PN and eccentricity order target.  The $s=+2$ code was then extended to the 10PN results
presented here.  The expansions also reproduce our earlier Schwarzschild results \cite{ForsEvanHopp16,MunnETC20,Munn20}
in the limit $a\to0$ and, at $e=0$, the circular-equatorial limit of our inclined spherical-orbit fluxes
\cite{CastETC25}, which were computed with a related but distinct code.  Within the derived analytical expressions,
we find select spin-dependent contributions that are resummable to closed-form expressions in $e$, as was the case
for some eccentricity enhancement functions in the Schwarzschild limit \cite{ForsEvanHopp16,MunnETC20,Munn20}.

We then compare the full expansions to numerical results for $a=0.9M$, finding relative errors of $\sim10^{-4}$
at a dimensionless semi-latus rectum $p\approx10$ for $e=0.1$, degrading to $\sim10^{-2}$ at $e=0.7$, and falling
as a power law in $p$.  The rapid improvement in accuracy at larger separations suggests that numerically evaluated
10PN fluxes may replace Teukolsky-code results in certain applications.  For example, for $a=0.9M$ they could do so
in adiabatic (0PA) inspiral calculations wherever their error is below the $\sim10^{-7}$ accuracy of current
interpolated flux grids, such as that of the FastEMRIWaveforms (FEW) modeling framework \cite{ChapBirdETC25}, which
holds for $p\gtrsim20$ at $e=0.1$, rising to $p\gtrsim75$ at $e=0.7$.

The paper has the following organization. Starting with Sec.~\ref{sec:background}, we discuss the source motion along
Kerr geodesics, the Teukolsky formalism \cite{Teuk73,Teuk72} used to compute the fluxes, and the analytic expansion
procedure \cite{KavaOtteWard16,Munn20,CastETC25,CastETC25b} used to obtain the PN expressions.  We then discuss the
results in Sec.~\ref{sec:results}, presenting first a general structure of the PN flux expressions and listing the
individual energy flux contributions up to $5$PN, with specific contributions, such as a $6.5$PN contribution and the
leading spin contributions up to $8$PN, highlighted for further discussion. A similar presentation for the angular
momentum fluxes is given in Appendix~\ref{app:AngMomFlux}.  A numerical comparison against flux results from an
independent Teukolsky code and an interpolated data set from FEW is presented and discussed in
Sec.~\ref{sec:numComp}.  Finally, in Sec.~\ref{sec:conclusion}, we provide a summary and outlook for future work.
Note here that we present select flux contributions to facilitate discussion; the complete fluxes through 10PN and
$e^{20}$ are publicly available through online repositories \cite{BHPTK18,UNCGrav22,PostNewtonianSelfForce}.

In this paper, we use the metric signature $(-,+,+,+)$ and, unless noted otherwise, we scale our units such that 
$c=G=1$.  For the MST formalism, we follow the notation used by Sasaki and Tagoshi \cite{SasaTago03} in their 
2010 revision.  For black hole perturbation theory more generally, we follow the conventions of \cite{PounWard22}. 

\section{Background}
\label{sec:background}

\subsection{Eccentric-equatorial geodesics in Kerr}
\label{sec:geoMotion}

In the absence of radiation reaction, the secondary is treated as a point mass $\mu$ in bound geodesic motion about a 
Kerr black hole of mass $M$ and spin $a = J/M$, with $\mu/M \ll 1$.  We use Boyer-Lindquist coordinates 
$x^\a = \{t,r,\th,\vp\}$ with the line element
\begin{align} 
    \begin{autobreak} 
    ds^2 = 
    -\l(1-\frac{2Mr}{\Sig}\r)dt^2
    -\frac{4aMr\sin^2\th}{\Sig}dtd\vp 
    +\frac{\Sig}{\D}dr^2
    +\Sig d\th^2
    +\l(\varpi^2+\frac{2Mra^2\sin^2\th}{\Sig}\r)\sin^2\theta d\vp^2 ,
    \end{autobreak}
\end{align}
and with 
\begin{align}
    \Sig &= r^2+a^2\cos^2\th, \\
    \D &= r^2-2Mr+a^2, \\
    \varpi &= \sqrt{r^2+a^2} .
\end{align}

In this paper, the motion $x^\a_p(\tau)$ of the point mass is confined to the equatorial plane, $\theta = \pi/2$, but 
is allowed to be eccentric. Stationarity and axisymmetry provide the conserved specific energy and axial angular 
momentum,
\begin{align}
\mathcal{E} = -\xi^\a_{(t)}u_\a=-u_t, \quad
\mathcal{L}_z = \xi^\a_{(\vp)}u_\a=u_\vp,
\end{align}
where $\xi^\a_{(t)}$ and $\xi^\a_{(\vp)}$ are the corresponding Killing vectors and $u^\a=dx_p^\a/d\tau$ is the 
particle four-velocity.  The Carter constant vanishes for equatorial motion.

At leading order in the mass ratio, the motion is governed by the Kerr geodesic equations \cite{Cart68,Schm02}, 
specialized to eccentric-equatorial orbits \cite{GlamKenn02}
\begin{align}
    \frac{dt_p(\tau)}{d\tau} &= \frac{\l(r_p \varpi^2_p +2Ma^2\r)\mathcal{E}-2Ma\mathcal{L}_z}{r_p\D_p}, \\
    \frac{dr_p(\tau)}{d\tau} &= \pm \frac{\sqrt{R(r_p)}}{r_p^2}, \\
    \frac{d\vp_p(\tau)}{d\tau} &= \frac{2Ma\mathcal{E}+\l(r_p-2M\r)\mathcal{L}_z}{r_p\D_p},
\end{align}
where
\begin{align}\label{Rdefinition}
    R(r) =& 
    \l( \varpi^2 \mathcal{E}-a\mathcal{L}_z\r)^2
    -\Delta\l(r^2+\l(\mathcal{L}_z-a\mathcal{E}\r)^2\r), \\
        =& (1-\mathcal{E}^2)(r_1-r)(r-r_2)(r-r_3)(r-r_4). \notag
\end{align}
Here, $r_1,r_2,r_3,r_4$ are the roots of $R(r)$ and the subscript $p$ represents functions evaluated at the location 
of the particle.  

For generic orbits, the coordinate equations depend on both $r_p$ and $\th_p$, requiring the use of Mino time $\la$
\cite{Mino03} to decouple the motion.  This problem is avoided in the equatorial plane, allowing the 
geodesic equations to be solved in a straightforward manner.  As in the Schwarzschild case, we reparameterize our 
equations with Darwin's geometric parameters
\begin{widetext}
\begin{align}
    r_p(\chi) = \frac{pM}{1+e\cos\chi},
\end{align}
where $p$ is the dimensionless semi-latus rectum, $e$ is the 
eccentricity, and $\chi$ is the relativistic anomaly \cite{Darw61,CutlKennPois94,BaraSago10}; we use these geometric 
parameters to facilitate later PN expansions.  Thus, for equatorial bound motion, $(p,e)$ provide an alternative 
parameterization to $(\mathcal{E},\mathcal{L}_z)$ and $1/p$ serves as a natural PN compactness parameter.  
In terms of $p$ and $e$, the roots of $R(r)$ simplify to 
\begin{align}
    r_1 = r_{\rm max} = \frac{pM}{1-e}, \qquad r_2 = r_{\rm min} = \frac{pM}{1+e}, 
\qquad r_3 = \frac{2M}{1-\mathcal{E}^2} - \frac{2pM}{1-e^2}, \qquad r_4 = 0.
\end{align}
Equating like powers of $r$ and using the definitions $q \equiv a/M$ and 
$\hat{\mathcal L} \equiv {\mathcal L}_z/M - q {\mathcal E}$, we can solve for $\mathcal E$ and $\hat {\mathcal{L}}$, 
yielding 
\begin{align}  \label{eq:geoELeqs}
\mathcal{E}^2 &= 1-\frac{\left(1-e^2\right)}{p} \Bigg[1- \frac{\left(1- e^2\right)}{\left(p-3-e^2\right)^2 p -4 q^2
   \left(1-e^2\right)^2} \bigg(q^2 \left(1+3 e^2+p\right) +p \left(p-3-e^2\right)  \notag 
\\ 
& \qq \qq \qq \qq \qq
   - (2q p^{-1/2} ) 
   \sqrt{q^4 \left(1-e^2\right)^2+\left((p - 2)^2 -4 e^2\right)  p ^2
+2 q^2 p \left(p - 2 +e^2 (2+p)\right)}\bigg)\Bigg] ,   \\
\frac{\hat{\mathcal{L}}^2}{p^2} &= \frac{q^2 (1+3 e^2+p)+  p \left(p-3-e^2\right)  - (2 q p^{-1/2}) 
\sqrt{q^4 (1-e^2)^2 + ((p-2)^2-4 e^2 )  p^2+2 q^2 p (p-2+e^2 (2+p) )}}{ (p-3-e^2)^2  p 
-4 q^2 (1-e^2)^2 }.
\end{align}
There is an issue of what convention to adopt for describing retrograde orbits.  It amounts to a choice of whether to 
flip the sign of the dimensionless spin parameter $q$ to be negative or to flip the sign of $\hat {\mathcal L}$.  In 
this paper, we adopt the convention that $\hat {\mathcal L} > 0$ (so that distant orbits always advance in the positive 
$\varphi$ direction) and prograde and retrograde correspond to $q>0$ or $q<0$, respectively.  There is no loss of 
generality with this assumption, since the coordinate system can always be oriented such that $\hat {\mathcal L} > 0$. 

Using this parameterization, one radial libration corresponds to a $2\pi$ increase in $\chi$.  Substituting 
$r_p(\chi)$ into the geodesic equations gives evolution equations for the remaining Boyer-Lindquist coordinates. 
Then, we can simplify the results using Eq.~\eqref{eq:geoELeqs} to get \cite{BiniDamoGera16c,BiniGera19a} 
\begin{align}
\label{eq:geoEqsChi}
\frac{d \tau_p}{d\chi} 
&= \frac{M}{v^3 (1+e \cos\chi )^2 \sqrt{1 + \hat{\mathcal{L}}^2 v^4 (e^2 - 2e \cos\chi - 3)}}  , \notag \\
\frac{dt_p}{d\chi} &= \left(\frac{d\tau_p}{d\chi}\right)  \frac{\mathcal{E} + \mathcal{E} q^2 v^4 (1+ e \cos \chi)^2 
- 2 q v^6 \hat{\mathcal{L}} (1+e\cos\chi)^3}{1 - 2 v^2 (1+ e\cos\chi) + q^2 v^4 (1+ e\cos\chi)^2  }  , \notag \\ 
\frac{d\vp_p}{d\chi} 
&= \frac{v [\hat{\mathcal{L}} + q {\mathcal E} - 2 v^2 \hat{\mathcal{L}} (1+ e \cos \chi)]}{[1 - 2 v^2 (1+ e\cos\chi) 
+ q^2 v^4 (1+ e\cos\chi)^2]\sqrt{1 + \hat{\mathcal{L}}^2 v^4 (e^2 - 2 e \cos \chi - 3) }  }  ,
\end{align}
where $v \equiv 1/\sqrt{p}$ is a PN velocity parameter.  The fundamental periods and frequencies 
then follow as 
\begin{align}
\label{eq:tandvpChi}
T_r &= \int_0^{2 \pi} \left( \frac{dt_p}{d\chi} \right) d\chi, \q  \O_r = \frac{2 \pi}{T_r}, \notag \\
\vp_p(\chi) 
&= \int_0^{\chi} \left( \frac{d\vp_p(\chi')}{d\chi'} \right) d\chi', \q  \O_\vp = \frac{\vp_p(2 \pi)}{T_r}  .
\end{align}
\end{widetext}
Analytical expressions for $\O_r$ and $\O_\vp$ have also been derived by Fujita and Hikida \cite{FujiHiki09}, but for 
our purposes, we find it more convenient to work directly with Eq.~\eqref{eq:geoEqsChi}, expanding the differential 
equations directly in $v$ and $e$ before integrating.

\subsection{The Teukolsky equation}
\label{sec:TeukRev}

As the small body moves along its orbit, it introduces a gravitational perturbation to the background Kerr 
spacetime.  The perturbation can be represented by either of two Weyl scalars, $\Psi_0$ and $\Psi_4$, satisfying 
the $s = +2$ and $s = -2$ Teukolsky equations, respectively \cite{Teuk72,Teuk73}, 
\begin{align}
    \mathcal{O}_{0}\Psi_0 &= 8\pi\mathcal{S}_{0}T, \\
    \mathcal{O}_{4}\Psi_4 &= 8\pi\mathcal{S}_{4}T.
\end{align}
Here $\mathcal{O}_{0,4}$ are Teukolsky operators and $\mathcal{S}_{0,4}$ are the respective decoupling operators 
acting on the stress-energy tensor $T_{\mu\nu}$ (\cite{PounWard22}, Secs.~4.1.2 and 4.1.7; 
\cite{WardKavaDola24}).  Following \cite{PounWard22}, the Weyl scalars are 
$\Psi_0 = C_{\a\b\g\d}\,l^\a m^\b l^\g m^\d$ and $\Psi_4 = C_{\a\b\g\d}\,n^\a \bar{m}^\b n^\g \bar{m}^\d$, where 
$(l^\a, n^\a, m^\a, \bar{m}^\a)$ is the Kinnersley tetrad \cite{Kinn69}, normalized so that $l^\a n_\a = -1$ and 
$m^\a \bar{m}_\a = 1$.  With these definitions, far from the source 
$\Psi_4 \sim -\tfrac{1}{2}\ddot{h}_{\bar{m}\bar{m}}$, where $h_{\bar{m}\bar{m}} = h_+ - i h_\times$ is the 
transverse-traceless strain \cite{PounWard22}, which easily connects to the fluxes at infinity.  Using the 
Kinnersley tetrad, we adopt the following separated mode decomposition for the Weyl scalars
\begin{widetext}
\begin{align}
    \zeta^{|s|-s}\Psi_{-s+2} &= \int_{-\infty}^{\infty} 
\sum_{\ell m}{}_{s}R_{\ell m\omega}(r){}_{s}S_{\ell m}(\theta;a\omega)e^{-i\omega t+im\varphi}d\omega, 
\label{eq:weylDec}
\end{align} 
where $\zeta = r - ia\cos\theta$.  In addition, the source functions have a similar mode decomposition
\begin{align}
    -16\pi\Sigma\zeta^{|s|-s}\mathcal{S}_{-s+2}T &= \int_{-\infty}^{\infty}\sum_{\ell m} 
{}_{s}T_{\ell m\omega}(r) {}_{s}S_{\ell m}(\theta;a\omega)e^{-i\omega t+im\varphi} d\omega. 
\label{eq:sourceDec} 
\end{align}
Note that the Fourier transform convention adopted here places the $1/(2\pi)$ factor on the inverse transform with the 
time integral.  The radial mode functions are ${}_{s}R_{\ell m\omega}(r)$ while the full spin-weighted spheroidal 
harmonic angular functions are written either as ${}_{s}S_{\ell m}(\theta;a\omega)e^{im\varphi}$ or 
${}_{s}S_{\ell m}(\theta, \varphi;a\omega)$.  These functions satisfy the separated radial Teukolsky and 
spin-weighted spheroidal harmonic equations, respectively,  
\begin{gather}
    \left[\Delta^{-s}\frac{d}{dr}\left(\Delta^{s+1}\frac{d}{dr}\right) 
+ \frac{K^2-2is(r-M)K}{\Delta}+4is\omega r-{}_{s}\lambda_{\ell m\omega}\right]{}_{s}R_{\ell m\omega} 
= {}_{s}T_{\ell m\omega}, \label{eq:radTEq} 
\\
    \left[\frac{1}{\sin\theta}\frac{d}{d\theta}\left(\sin\theta\frac{d}{d\theta}\right)
-a^2\omega^2\sin^2\theta-\frac{(m+s\cos\theta)^2}{\sin^2\theta}-2a\omega s\cos\theta+s
+2am\omega+{}_{s}\lambda_{\ell m\omega}\right]{}_{s}S_{\ell m} = 0, \label{eq:sphTEq}
\end{gather}
\end{widetext}
where $K = (r^2+a^2)\omega - am$ and the separation constant ${}_{s}\lambda_{\ell m\omega}$ depends on $s, \ell, m$ 
and $a\omega$.  An alternative common choice of separation constant is 
${}_{s}A_{\ell m\omega} = {}_{s}\lambda_{\ell m\omega} + 2am\omega - a^2\omega^2$ \cite{Vand16,PounWard22}.  In this 
work, we adopt the standard convention \cite{PounWard22} for orthonormality of the spin-weighted spheroidal harmonics 
\begin{align}
\int d\Omega \ {}_{s}S_{\ell m}(\theta;a\omega) e^{im\varphi} \, 
{}_{s}S_{\ell^\prime m^\prime}^*(\theta;a\omega) e^{-i m^\prime \varphi} 
= \delta_{\ell \ell^\prime} \delta_{m m^\prime} .
\end{align}
For the homogeneous radial Teukolsky functions, we choose the ``in'' solution to be purely ingoing at the horizon and 
the ``up'' solution to be purely outgoing at (future) null infinity, which correspond to the transmitted waves in 
their respective scattering problems \cite{Galt82,PounWard22}.  We normalize these homogeneous solutions to have unit 
transmission amplitudes, leading to the following asymptotic behavior
\cite{Teuk73}
\begin{align}
    &{}_{s}R_{\ell m \omega}^\text{in} \sim \Delta^{-s}e^{-i(\omega- m\Omega_H)r_*} \quad
    &{\rm as} \quad r \to r_+ ,
\label{eq:boundin} 
\\
    &{}_{s}R_{\ell m \omega}^\text{up} \sim r^{-2s-1}e^{i\omega r_*}  \quad
    &{\rm as} \quad r \to \infty . 
\label{eq:boundup} 
\end{align}
Here $\Omega_H = a/(2 M r_{+}) = a/(2M(M + \sqrt{M^2-a^2}))$ is the horizon angular velocity and the tortoise 
coordinate $r_*$ is defined from $dr_*/dr = (r^2+a^2)/\Delta$.  We fix its integration constant as in
Eq.~(21) of \cite{SasaTago03},
\begin{align}
    r_* = r + \frac{2Mr_+}{r_+-r_-}\log\frac{r-r_+}{2M} - \frac{2Mr_-}{r_+-r_-}\log\frac{r-r_-}{2M} ,
\label{eq:rstar}
\end{align}
where $r_\pm = M \pm \sqrt{M^2-a^2}$ are the outer and inner horizons.  The phases of the amplitudes 
${}_{s}Z^\text{in/up}_{\ell m\omega}$ (defined below) depend on this choice; the fluxes do not. 

For a bound orbit, the continuous frequency spectrum in Eqs.~\eqref{eq:weylDec}--\eqref{eq:sourceDec} is converted 
into a discrete frequency spectrum due to the periodicity in the motion \cite{DrasHugh04}.  For eccentric-equatorial 
orbits in particular, the discrete frequencies are
\begin{align}
\label{eq:omegamn}
    \omega_{mn} = m\Omega_\varphi + n\Omega_r,
\end{align}
where $n$ is an integer representing the harmonics of radial libration, and the Fourier integrals over $\omega$ become 
sums over $n$.  Accordingly, we replace the continuous-frequency label $\ell m\omega$ by the discrete label 
$\ell mn$, e.g., ${}_{s}R_{\ell m\omega} \to {}_{s}R_{\ell mn}$, ${}_{s}T_{\ell m\omega} \to {}_{s}T_{\ell mn}$, 
and ${}_{s}S_{\ell m}(\theta,\varphi;a\omega) \to {}_{s}S_{\ell mn}$.  

\subsection{Particular solution of the Teukolsky equation with a point-particle source}
\label{sec:inhomTeuk}

To obtain the particular solution of the Teukolsky equation, we use a Green's function method
\begin{align}
    {}_{s}R_{\ell m \omega}(r) =& \int_{r_+}^\infty dr'{}_{s}
    \mathcal{G}_{\ell m \omega}(r,r') \Delta^{s}(r') {}_{s}T_{\ell m \omega}(r'),
\end{align}
where ${}_{s}\mathcal{G}_{\ell m \omega}(r,r')$ is the Green's function for the radial Teukolsky equation.  The 
source amplitudes ${}_{s}T_{\ell m \omega}(r')$ arise from tetrad projections of the stress-energy tensor, which we 
take to be a point source confined to eccentric-equatorial motion
\begin{align}
\label{eq:stressT}
    T_{\mu\nu} = \mu\frac{u_\mu u_\nu}{\Sigma u^t}\delta(r-r_p)\delta(\theta-\pi/2)\delta(\vp-\vp_p) .
\end{align}
Here, $u_\mu = g_{\mu\nu}u^\nu$ and $u^\mu = dx_p^\mu/d\tau$ is the particle four-velocity.  While details of the 
calculation of ${}_{s}T_{\ell m \omega}(r')$ from $T_{\mu\nu}$ can be found in
\cite{Teuk73,MinoETC97,SasaTago03,SrivChen21,Nasi25}, we note that the projected source 
terms, prior to transfer to the frequency domain, are distributional with 
$\delta(r - r_p(t))$, $\delta^\prime(r - r_p(t))$, and $\delta^{\prime\prime}(r - r_p(t))$ dependence.

With Eqs.~\eqref{eq:boundin} and \eqref{eq:boundup} as the boundary conditions, ${}_{s}\mathcal{G}_{\ell m\omega}$ 
has the following structure, 
\begin{multline}
    {}_{s}\mathcal{G}_{\ell m \omega}(r,r') 
= \frac{1}{{}_{s}W_{\ell m\omega}}
\bigg[{}_{s}R_{\ell m\omega}^\text{up}(r'){}_{s}R_{\ell m \omega}^\text{in}(r)\Theta(r'-r) \\ 
    + {}_{s}R_{\ell m \omega}^\text{in}(r') {}_{s}R_{\ell m\omega}^\text{up}(r)\Theta(r-r')\bigg],
\end{multline}
where ${}_{s}W_{\ell m \omega}$ is the invariant Wronskian, a constant computed from 
\begin{multline}
    \frac{{}_{s}W_{\ell m \omega}}{\Delta^{s+1}} 
= {}_{s}R_{\ell m\omega}^\text{in} \frac{d{}_{s}R_{\ell m\omega}^\text{up}}{dr} 
- {}_{s}R_{\ell m\omega}^\text{up}\frac{d{}_{s}R_{\ell m\omega}^\text{in}}{dr}.
\end{multline}

For bound motion, $r_{\rm min} \le r_p \le r_{\rm max}$, integration over the source leads to vacuum regions 
on either side in which the particular solution becomes
\begin{align}
    {}_{s}R_{\ell m \o}(r) = \begin{cases}
        {}_{s}Z_{\ell m \o}^\text{in} \, {}_{s}R_{\ell m \o}^\text{in}(r), & r<r_{\rm min} \\
        {}_{s}Z_{\ell m \o}^\text{up} \, {}_{s}R_{\ell m \o}^\text{up}(r), & r>r_{\rm max} 
    \end{cases},
\end{align}
where ${}_{s}Z_{\ell m \o}^\text{up}$ and ${}_{s}Z_{\ell m \o}^\text{in}$ are asymptotic amplitudes given by
\begin{align}
    {}_{s}Z_{\ell m \o}^\text{in/up} = \int_{r_{\rm min}-\eta}^{r_{\rm max}+\eta} dr' 
\frac{\Delta^{s}(r') \, {}_{s}R_{\ell m \o}^\text{up/in}(r')}{{}_{s}W_{\ell m \o}}{}_{s}T_{\ell m \o}(r') , 
\label{eq:modefunctions}
\end{align} 
and where $\eta > 0$ is small enough to ensure the lower bound of integration is outside the outer horizon but is 
otherwise arbitrary.  This adjustment to the range of integration is a necessary caveat to this 
otherwise standard procedure.  It is required due to the distributional nature of the source function in 
Eq.~\eqref{eq:sourceDec}, with the first and second derivatives of $\delta(r - r_p(t))$ producing singular behavior in 
${}_{s}T_{\ell m \o}(r')$ at the end points.  However, in practice our code reverses the order 
of integration, eliminating this issue.  For a fixed $\omega$, the code integrates the space- and time-dependent 
source in Eq.~\eqref{eq:sourceDec} times a particular kernel over $r$, $\theta$, and $\vp$ first, including 
integrations by parts in $r$ and $\theta$, leaving a smooth periodic function, written in terms of $\chi$.  The smooth 
periodic function in $\chi$ for each mode is then integrated over one radial libration, resulting in the value 
of the discrete amplitudes ${}_{s}Z_{\ell m n}^\text{in/up}$, connected to the continuous-in-$\omega$ amplitudes 
via the Dirac comb, ${}_{s}Z_{\ell m\o}^\text{in/up} = \sum_n {}_{s}Z_{\ell mn}^\text{in/up}\,\delta(\o-\o_{mn})$.  
This procedure, reversing the order of integration, is described in \cite{HoppEvan10} for the simpler 
Regge-Wheeler-Zerilli problem in Schwarzschild.

For computing the fluxes at infinity, the outgoing asymptotic amplitudes, ${}_{s}Z_{\ell m \o}^\text{up}$, are
used to find individual mode contributions to the fluxes.  Switching from $\ell m \o$ to discrete harmonics 
$\ell m n$, the energy and angular momentum fluxes are computed \cite{TeukPres74,MinoETC97,GlamKenn02,DrasHugh06} 
from the incoherent sums 
\begin{align}
\left\langle\frac{dE}{dt}\right\rangle_{\infty} 
&=\frac{1}{4\pi}\sum_{\ell m n}\frac{\left|{}_{-2}Z_{\ell m n}^\text{up}\right|^2}{\o_{mn}^2}, 
\label{eq:fluxEn1} \\
&=\frac{4}{\pi} 
\sum_{\ell m n}\frac{\o_{mn}^6}{|C^\text{TS}_{\ell m n}|^2}\left|{}_{2}Z_{\ell m n}^\text{up}\right|^2 ,
\label{eq:fluxEn2}   
\end{align}
and 
\begin{align}
\left\langle\frac{dL_z}{dt}\right\rangle_\infty 
&=\frac{1}{4\pi}\sum_{\ell m n}\frac{m\left|{}_{-2}Z_{\ell m n}^\text{up}\right|^2}{\o_{mn}^3}, 
\label{eq:fluxLz1} \\
&=\frac{4}{\pi}
\sum_{\ell m n}\frac{m\o_{mn}^5}{|C^\text{TS}_{\ell m n}|^2}\left|{}_{2}Z_{\ell m n}^\text{up}\right|^2,
\label{eq:fluxLz2}
\end{align}
where $C^\text{TS}_{\ell m n}$ is the Teukolsky--Starobinsky constant \cite{TeukPres74,StarChur74},
\begin{equation}
\begin{split}
    |C^\text{TS}_{\ell m n}|^2 ={}& \l[({}_{-2}\la_{\ell m n}+2)^2 + 4am\o_{mn} - 4a^2\o_{mn}^2\r] \\
    &\times\l({}_{-2}\la_{\ell m n}^2 + 36am\o_{mn} - 36a^2\o_{mn}^2\r) \\
    &+ (2\,{}_{-2}\la_{\ell m n}+3)\l(96a^2\o_{mn}^2 - 48am\o_{mn}\r) \\
    &+ 144\o_{mn}^2\l(M^2 - a^2\r),
\end{split}
\label{eq:CTS}
\end{equation}
written with the $s=-2$ separation constant ${}_{-2}\la_{\ell m n}$ of Eq.~\eqref{eq:radTEq}, evaluated at 
$\o = \o_{mn}$.  Finally, the horizon fluxes are computed using the ingoing asymptotic amplitudes 
${}_{s}Z_{\ell m n}^\text{in}$ in the associated horizon flux formulas \cite{TeukPres74}.  The high-order PN 
expansions of the horizon fluxes will be the subject of an upcoming paper. 

\subsection{Post-Newtonian expansions}
\label{sec:PNExpan}

This section addresses the PN expansion of the spin-weighted spheroidal harmonics and the
radial Teukolsky mode functions, which are steps toward obtaining PN expansions of the infinity-side fluxes.

\subsubsection{Separation eigenvalue and angular functions}

The first step in this process is to obtain the separation eigenvalue, ${}_{s}\la_{\ell m n}$, as a PN expansion.  
The spin-weighted spheroidal eigenvalue routine in the \texttt{SpinWeightedSpheroidalHarmonics} package of the
Black Hole Perturbation Toolkit \cite{BHPTK18,SWSH} is called and the result is expanded as a series in 
$\gamma = a\omega$ to adequately high order.  From there, it is converted to a series in $v$ by relating $\omega$ to 
$\omega_{mn}$ of the source motion [Eq.~\eqref{eq:omegamn}] and using PN series for $\Omega_\vp$ and $\Omega_r$.

We turn next to the angular functions.  As shown in Sec.~\ref{sec:TeukRev}, in the frequency domain and with 
axisymmetry, the angular dependence separates off into spin-weighted spheroidal harmonics 
${}_{s}S_{\ell m}(\th,\vp;a\o) = {}_{s}S_{\ell m}(\th;a\o)e^{im\vp}$.  To solve for any particular mode, we need 
to cast the time-dependent stress-energy tensor [Eq.~\eqref{eq:stressT}] into the frequency domain and then project 
its components with the tetrad and angular functions to obtain the source amplitudes ${}_{s}T_{\ell m\omega}(r)$.  
Because the Teukolsky equation does not separate in $r$ and $\theta$ except in the frequency domain, the order of 
these steps matters.

The azimuthal angle itself separates merely from axisymmetry.  When the source is integrated, the point-source 
behavior leads to a term of the form $\exp\{-im\vp_p(\chi)\}$.  The azimuthal coordinate advances with a mean rate 
$\O_\vp/\O_r$ in $\chi$ along with a modulation that is periodic in the radial libration, as seen by 
considering the behavior of Eq.~\eqref{eq:tandvpChi}.  The evolution can be PN expanded to yield
\begin{widetext}
\begin{align}
    e^{-im\vp_p(\chi)} &= e^{-im(\O_\vp/\O_r)\chi}\Big\{1-\frac{ime\sin\chi}{p} 
    -\frac{im}{p^2}\left[e(9-q^2)\sin\chi+\frac{e^2(3-2q^2)}{8}\sin 2\chi\right] \notag\\
    &\quad -\frac{m^2e^2\sin^2\chi}{2p^2}+\frac{20imeq\sin\chi}{p^{5/2}}+O\left(\frac{1}{p^{3}}\right)\Big\}, \\
    \frac{\O_\vp}{\O_r} &= 1+\frac{3}{p}-\frac{4q}{p^{3/2}} 
    +\frac{3(18+e^2+2q^2)}{4p^2}-\frac{36q}{p^{5/2}}+O\left(\frac{1}{p^{3}}\right) , \notag
\end{align}
a form that is amenable to periodic source-term integration.
\end{widetext}

We next consider the polar part of the angular harmonics.  In general terms, we can use the small-frequency 
expansion of the spin-weighted spheroidal harmonics in terms of spin-weighted spherical harmonics 
\cite{BreuRyanWall77,KavaOtteWard16}
\begin{align}
    {}_{s}S_{\ell m}(\th,\vp;a\o) = \sum_{j=j_{\rm min}}^{\infty} b^{\ell}_{jm}(a\o) {}_{s}Y_{jm}(\th,\vp) ,
\end{align}
where $j_{\rm min} = \max(|s|,|m|)$, as ${}_{s}Y_{jm}$ exists only for $j \ge |s|$ and $|m| \le j$.  The same 
bound holds for the mode index, $\ell \ge j_{\rm min}$.  (This expansion does not involve the azimuthal 
factor $e^{im\vp}$.)  The coefficients $b^\ell_{jm}(a\o)$ for given harmonic numbers $\ell$ and 
$m$ ($s$ suppressed) form a potentially infinite set over $j$.  In practice, however, the coefficients fall off 
exponentially as $j$ moves away from $\ell$ in either direction, with 
$b^\ell_{jm}(a\o) \sim (a\o)^{|\ell-j|}$.\footnote{See \cite{Warb15} for a scalar ($s=0$) case example.}  
Adopting this as the basis for a series expansion, the projection of the angular functions on the spin-weighted 
spherical harmonic basis becomes
\begin{align}
    {}_{s}S_{\ell m}(\th,\vp;a\o) 
    = \sum_{k = 0}^{\infty}(a\o)^k\left[\sum_{i=-k}^{k}d_{ik}^{\ell m}{}_{s}Y_{\ell+i,m}(\th,\vp)\right] ,
\end{align}
where terms with $\ell + i < j_{\rm min}$ are absent.  Powers of $a \omega$ again relate to powers in the PN 
expansion.  The coefficients $d_{ik}^{\ell m}$ and the resulting expansion of ${}_{s}S_{\ell m}(\th,\vp;a\o)$ are 
computed in the \texttt{SpinWeightedSpheroidalHarmonics} 
package of the Black Hole Perturbation Toolkit \cite{BHPTK18,SWSH}.  With the series expansion in this form, it is 
easy to calculate $\theta$-derivatives of the spheroidal harmonics.  The PN expansion of 
${}_{s}S_{\ell m}(\th,\vp;a\o)$ and the first two derivatives in $\theta$ are then specialized to $\theta = \pi/2$ 
as part of the source integration.

\subsubsection{Radial Teukolsky function}

We construct homogeneous solutions to the radial Teukolsky equation, Eq.~\eqref{eq:radTEq}, through the MST method 
\cite{ManoSuzuTaka96a,ManoSuzuTaka96b}.  In the MST method, the solutions are represented by a convergent sum of 
special functions.  There are multiple representations known, and we specifically use the Coulomb wave function 
representation \cite{Leav86,ManoSuzuTaka96a,ManoSuzuTaka96b,SasaTago03} (see \cite{YostWheeBrei36} for use in 
quantum scattering).  Detailed discussions of this representation are given in Secs.~4.3--4.4 of \cite{SasaTago03}, 
and in Appendix B of \cite{Thro10}.

\begin{widetext}
The Coulomb wave representation is constructed from the solutions\footnote{The mode indices are suppressed.} 
$R_C^\nu$ to the Teukolsky equation \cite{SasaTago03,Leav86}, 
\begin{multline}
    R_{C}^\nu = \left(1-\frac{\eps \ka}{z}\right)^{-s-i(\eps+\tau)/2} e^{-iz}2^{s}(2z)^{\nu-s} 
    \frac{\G(\nu+1-s+i\eps)}{\G(\nu+1+s-i\eps)}\times \\
    \sum_{j=-\infty}^{\infty}a_{j}^{\nu} (-2i z)^{j}\frac{\G(j+\nu+1+s-i\eps)}{\G(2j+2\nu+2)} 
    \Phi(j+\nu+1-s+i\eps,2j+2\nu+2;2iz). 
\label{eq:RCSum}
\end{multline}
Here $z = \o(r-r_-)$, $\eps = 2M\o$, $\tau = (\eps-mq)/\ka$, $\ka = \sqrt{1-q^2}$, and 
$\Phi(\alpha,\beta;z)$ is the regular confluent hypergeometric function.  The parameter $\nu$ is the renormalized 
angular momentum and $a_j^{\nu}$ are the MST coefficients, determined by a three-term recurrence relation 
(within which ${}_{s}\la_{\ell m n}$ appears).  The parameter $\nu$ satisfies a constraint equation that guarantees 
falloff in the $a_j^{\nu}$ and the convergence of the series as $j \to \pm \infty$ (i.e., a minimal solution).  
In using the MST method, we normalize the coefficients by $a_0^{\nu} = 1$; those of the second solution are 
$a_j^{-\nu-1} = a_{-j}^{\nu}$ (Sec.~4.2 of \cite{SasaTago03}).  The result is that these solutions will have in 
general non-unit transmission amplitudes (see below).  The solutions $R_C^{\nu}$ and $R_C^{-\nu-1}$ form a 
pair of linearly independent homogeneous solutions, and they 
are related to the ``in'' and ``up'' solutions by 
\begin{align}
    {}_{s}B_{\ell m \o}^\text{trans}{}_{s}R_{\ell m \omega}^{\mathrm{in}} 
&= K_{\nu}R_C^{\nu} + K_{-\nu-1}R_{C}^{-\nu-1}, \label{eq:rInCWR} \\
    {}_{s}C_{\ell m \o}^\text{trans}{}_{s}R_{\ell m \o}^{\mathrm{up}} 
&= e^{-\pi\eps-i\pi s}\frac{\sin(\pi(\nu+s-i\eps))}{i\sin(2\pi\nu)}
\left[R_C^{-\nu-1}+ie^{-i\pi\nu}\frac{\sin(\pi(\nu-s+i\eps))}{\sin(\pi(\nu+s-i\eps))}R_C^\nu\right], 
\label{eq:rUpCWR}
\end{align}
where the factor $K_\nu$ is dependent on $\o$ but not on $r$, and ${}_{s}B_{\ell m \o}^\text{trans}$ and 
${}_{s}C_{\ell m \o}^\text{trans}$ are asymptotic constants extracted from the MST solutions.  Complete expressions 
are given in Eqs.~(165), (167), and (170) of \cite{SasaTago03}.  In addition, we adopt a factorization that has 
proven useful in previous applications \cite{CastETC25,CastETC25b}, defining 
${}_{s}\hat{R}^\text{in}_{\ell m\o} \equiv {}_{s}R^\text{in}_{\ell m\o}/K_\nu$, so that 
Eq.~\eqref{eq:rInCWR} becomes \cite{NeefKavaOtte26}
\end{widetext}
\begin{align}
    {}_{s}B_{\ell m \o}^\text{trans}{}_{s}\hat{R}_{\ell m \omega}^\text{in} 
    = R_{C}^\nu + \frac{K_{-\nu-1}}{K_\nu}R_{C}^{-\nu-1} .
\end{align}
Since $K_\nu$ is independent of $r$, this rescaling leaves the amplitudes ${}_{s}Z^\text{up}_{\ell m\o}$ of
Eq.~\eqref{eq:modefunctions} unchanged, with the factor canceling between ${}_{s}R^\text{in}_{\ell m\o}$ and the
Wronskian ${}_{s}W_{\ell m\o}$.  However, the rescaled radial function is significantly easier to expand to high PN 
orders, as the ratio $K_{-\nu-1}/K_{\nu}$, also known as the ``tidal response function'' \cite{BautETC24}, has a 
simpler form due to cancellations.  More importantly, the ratio scales as $\omega^{2\ell-1}$ (as $\omega^{2\ell}$ 
when $m q = 0$), suppressing $R_{C}^{-\nu-1}$ with respect to $R_C^\nu$.

Expanding the MST solution into a high PN-order series proceeds in three steps,
like those of a numerical calculation.  First, a high PN-order expansion of the renormalized angular momentum, 
$\nu = \nu(s,\ell,m,a,\o)$, is calculated from a continued fraction, terminated at a specified depth determined by 
the desired PN order.  Second, the MST series coefficients, $a_j^\nu$, are PN-expanded to the desired order using 
the previously computed PN series for $\nu$.  Finally, the MST sum in Eq.~\eqref{eq:RCSum} and the mode functions 
in Eqs.~\eqref{eq:rInCWR}--\eqref{eq:rUpCWR} are assembled as high-order PN series.  Further details on this 
procedure are found in \cite{KavaOtteWard15,KavaOtteWard16,Munn23}.   

\subsubsection{Sum over modes}

In computing the fluxes, the sums in Eqs.~\eqref{eq:fluxEn1}--\eqref{eq:fluxLz2} formally have an infinite number 
of terms in both $\ell$ and $n$.  
In practice, however, both sums may be truncated to a finite number of terms.  The sum in $\ell$ may be related to 
the leading-order behavior of the radial functions, which are $\ell$-dependent 
\cite{KavaOtteWard15,KavaOtteWard16,Munn20}.  The flux at a given target PN order is then determined by a partial 
sum up to a finite $\ell_\text{max}$, since a mode of multipole order $\ell$ first contributes at relative 
$(\ell-2)$PN order.  For our 10PN results, this means summing through $\ell_\text{max} = 12$.  
Summing $\ell$ beyond that point merely affects the truncation error.  The sum over $n$, on the other hand, is 
related to the eccentricity dependence, with the asymptotic amplitudes exhibiting a power series in $e$ and with a 
leading-order term in $e$ that depends on $n$ as \cite{MunnETC20,Munn20}
\begin{align}
    {}_{s}Z_{\ell m n}^\text{up} \sim e^{|n|} .
\end{align}
The squared modulus of an amplitude then yields a per-mode flux contribution that has a leading eccentricity 
dependence of $e^{2|n|}$.  The flux falls off strongly with higher-order harmonics of the radial libration 
frequency.  For a graphical representation of the falloff in energy flux contributions with $n$, see 
Fig.~5 of \cite{ForsEvanHopp16}.  Since we truncate our eccentricity expansions in this paper at order $e^{20}$, 
there is a beneficial limit on the number of $n$ modes we need to compute---in this case 21 $n$ modes.  In general,
for flux expansions accurate up to $e^N$, the range of $n$ modes 
required is
\begin{align}
    -\lfloor N/2 \rfloor \leq n \leq \lfloor N/2 \rfloor .
\end{align}
This result is independent of $\ell$, $m$, and the target PN order.  

\begin{widetext}
\section{Results}
\label{sec:results}

\subsection{Structure}

The general structure of the PN-expanded infinity-side gravitational fluxes has been previously elucidated 
\cite{Blan14,Fuji12a,Fuji12b,Munn20}. We start with the infinity-side gravitational energy flux, which has the form 
\begin{align}
\begin{autobreak}
    \l\langle \frac{dE}{dt} \r\rangle_\oo = 
    \frac{32}{5}\frac{\mu^2}{M^2}\left(1-e^2\right)^{3/2}p^{-5}\bigg[ 
    \mathcal{A}_0
    +p^{-1} \mathcal{A}_1
    +p^{-3/2} \mathcal{A}_{3/2}
    +p^{-2} \mathcal{A}_2
    +p^{-5/2} \mathcal{A}_{5/2}
    +p^{-3}\left[ \mathcal{A}_{3}+\log(p)\mathcal{A}_{3L}\right]
    +p^{-7/2}\mathcal{A}_{7/2}
    +p^{-4}\left[  \mathcal{A}_{4} +\log(p)\mathcal{A}_{4L}\right]
    +p^{-9/2}\left[  \mathcal{A}_{9/2} +\log(p)\mathcal{A}_{9/2L}\right] 
    +p^{-5} \left[ \mathcal{A}_{5}+\log(p)\mathcal{A}_{5L}\right] 
    +p^{-11/2} \left[ \mathcal{A}_{11/2}+\log(p)\mathcal{A}_{11/2L}\right] 
    +p^{-6} \left[ \mathcal{A}_6 + \log(p)\mathcal{A}_{6L}+\log^2(p) \mathcal{A}_{6L2} \right]
    +\cdots\bigg] \ . 
\end{autobreak}
\label{expansionfordEdt}
\end{align}
Here we use the same conventions as in \cite{CastETC25,CastETC25b} for the coefficients $\mathcal{A}_{iLj}$, which 
represent functions of $a$ and $e$.  The tag $i$ represents the relative PN order of the term, which takes integer 
or half-integer values, while the tag $Lj$ appears when a $\log^j p$ term is present.  For simplicity, we employ 
the following abbreviations $\mathcal{A}_{iL0} = \mathcal{A}_{i}$ and $\mathcal{A}_{iL1} = \mathcal{A}_{iL}$.  
Next, the general expression for the infinity-side angular momentum flux has a similar form 
\begin{align}
    \begin{autobreak}
    \left\langle\frac{dL_z}{dt}\right\rangle_{\oo} = 
        \frac{32}{5}\frac{\mu^2}{M}\left(1-e^2\right)^{3/2}p^{-7/2} \bigg[
        \mathcal{C}_0 
        +\mathcal{C}_{1}p^{-1}
        +\mathcal{C}_{3/2}p^{-3/2}
        +\mathcal{C}_{2}p^{-2}
        +\mathcal{C}_{5/2}p^{-5/2}
        +\left(\mathcal{C}_{3}+\mathcal{C}_{3L}\log(p)\right)p^{-3}
        +\mathcal{C}_{7/2}p^{-7/2} 
        +\left(\mathcal{C}_{4}+\mathcal{C}_{4L}\log(p)\right)p^{-4}
        +\left(\mathcal{C}_{9/2}+\mathcal{C}_{9/2L}\log(p)\right)p^{-9/2}
        +\left(\mathcal{C}_{5}+\mathcal{C}_{5L}\log(p)\right)p^{-5} 
        +\left(\mathcal{C}_{11/2}+\mathcal{C}_{11/2L}\log(p)\right)p^{-11/2}
        +\left(\mathcal{C}_{6}+\mathcal{C}_{6L}\log(p)+\mathcal{C}_{6L2}\log^2(p)\right)p^{-6} 
        +\cdots\bigg],
    \end{autobreak}
\end{align}
with the coefficients $\mathcal{C}_{iLj}$ also representing functions of $a$ and $e$. 

\end{widetext}

In the following subsections, we focus on the energy flux contributions and relegate the angular momentum flux 
contributions to Appendix \ref{app:AngMomFlux}.  Furthermore, since previous works 
\cite{ForsEvanHopp16,MunnEvan19a,MunnETC20,Munn20,MunnEvan20a} have investigated the detailed structure of flux 
terms in the Schwarzschild limit ($a = 0$), we introduce a decomposition of the flux components that focuses on 
spin-dependent terms
\begin{align}
    \mathcal{A}_{iLj} = \mathcal{A}_{iLj}^{(0)}(e) + \sum_{k\ge0} \mathcal{A}_{iLj}^{Sk}(a,e) .
\end{align}
Here $\mathcal{A}_{iLj}^{(0)}(e)$ represents the Schwarzschild limit with the terms in the sum carrying all of the 
spin dependence.  The superscript $Sk$ labels terms with an explicit power-law factor $q^k$, and the argument $a$ 
in these terms indicates that these coefficients may depend on $a$ also through $\kappa$, $\log\kappa$ and the 
functions $\Psi^{(r,\sigma)}$, $\bar\Psi^{(r,\sigma)}$, $\tilde\Psi^{(r,\sigma)}$ derived from 
polygamma functions, as defined below.  All of the terms in the sum, including $\mathcal{A}_{iLj}^{S0}$, vanish for 
$a=0$.  Thus, our presentation in this paper is restricted to new, spin-dependent terms found in the non-zero elements 
of the set $\mathcal{A}_{iLj}^{Sk}$.  However, because of their increasing complexity with 
PN order, we only display here the complete set of 
spin-dependent terms up to 5PN order.  Beyond 5PN, we display only select terms.  One such is the $S1$ term at 
6.5PN order, which illustrates the beginning of more complicated, non-polynomial spin dependence.  
We then give the eight leading-spin terms of the energy flux through 8PN (and, in Appendix~\ref{app:leadq}, those
of the angular momentum flux through 10PN), six of which have closed-form expressions in $e$.  Beyond these, 
the full flux component expressions are too cumbersome at higher PN order, and we relegate the entirety of the 
10PN flux expressions to online repositories \cite{BHPTK18,UNCGrav22,PostNewtonianSelfForce}.  

\subsection{Energy flux contributions}

We begin by noting the leading-order contribution, well known from the results of Peters and Mathews \cite{PeteMath63} 
\begin{align}
    \mathcal{A}_{0} = 1+\frac{73}{24}e^2+\frac{37}{96}e^4,
\end{align}
which reappears in some of the higher-order contributions.  The lowest-order spin contribution begins 
at 1.5PN, with the following expression 
\begin{align}
    \mathcal{A}_{3/2}^{S1} = -q\left(\frac{73}{12}+\frac{823}{24}e^2+\frac{949}{32}e^4+\frac{491}{192}e^6\right),
\end{align}
which is exact, terminating at $e^6$.  It is followed by similarly exact expressions at 2PN and 2.5PN
\begin{align}
    \mathcal{A}_{2}^{S2} &= q^2\left(\frac{33}{16}+\frac{359}{32}e^2+\frac{1465}{128}e^4+\frac{883}{768}e^6\right), \\
    \mathcal{A}_{5/2}^{S1} &= -q\bigg(\frac{283}{336}-\frac{723}{56}e^2+\frac{49883}{1344}e^4+\frac{53045}{448}e^6 
\notag \\
    &+\frac{205739}{14336}e^8-12\left(1-e^2\right)^{3/2}\mathcal{A}_{0}\bigg).
\end{align} 
Here $\mathcal{A}_{5/2}^{S1}$ was originally a series expansion in $e$, truncated to $e^{20}$.  The series expansion 
turned out to be resummable, resulting in an exact expression in $e$, similar in form to low PN contributions for 
the non-spinning fluxes.  We identify such closed forms by fitting the series through $e^{20}$ with functions of 
the form $P(e^2)+(1-e^2)^{1/2}Q(e^2)$ or $P(e^2)+(1-e^2)^{3/2}Q(e^2)$, with $P$ and $Q$ polynomials.  A fit is 
accepted as a closed form only if it reproduces every computed coefficient in the series; the forms found have 
between one and eight coefficients to spare.  Where the margin is a single over-determined coefficient, we regard 
that fit as still almost certainly, but not provably, exact based upon prior experience with longer expansions in the 
Schwarzschild limit \cite{ForsEvanHopp16,MunnETC20,Munn20}.

We then proceed to the 3PN flux contributions, which have two non-zero coefficients due to spin.  We list first the 
spin contribution with an exact expression 
\begin{align}
    \mathcal{A}_{3}^{S2} 
&= q^2\bigg(\frac{5267}{168} + \frac{162209}{672}e^2 + \frac{908003}{2688}e^4 + \frac{1050307}{5376}e^6 \notag \\
    & + \frac{718387}{43008}e^8 - 11\left(1-e^2\right)^{3/2}\mathcal{A}_{0}\bigg),
\end{align}
then the spin contribution with a non-terminating series in eccentricity 
\begin{align}
    \begin{autobreak}
        \MoveEqLeft
        \mathcal{A}_{3}^{S1} = 
        q\pi\bigg(-\frac{169}{6}
            -\frac{4339}{16}e^2
            -\frac{42271}{96}e^4
            -\frac{4867907}{27648}e^6
            -\frac{134279}{16384}e^8
            -\frac{2289029}{17694720}e^{10}
            +\frac{4322471}{132710400}e^{12}
            -\frac{68755189}{66588770304}e^{14}
            +\frac{13941221843}{106542032486400}e^{16}
            +\frac{19785821383121}{207117711153561600}e^{18}
            +\frac{2034085304147}{138078474102374400}e^{20}
        +\cdots\bigg),
    \end{autobreak}
\end{align}
computed to $e^{20}$.  Like the lowest-order tail term $\mathcal{A}_{3/2}^{(0)}$ \cite{BlanScha93,ForsEvanHopp16}, 
$\mathcal{A}_{3}^{S1}$ carries a factor of $\pi$ and is a non-terminating series in $e$: it is the leading tail 
contribution linear in spin, appearing 1.5PN orders above the nonspinning tail.

For the 3.5PN flux contributions, there are three non-zero coefficients with spin.  Two of them are exact 
expressions, which we present first below 
\begin{align}
    \begin{autobreak}
        \MoveEqLeft
        \mathcal{A}_{7/2}^{S1} = q\bigg[
            -\frac{130765}{9072}
        +\frac{7379371}{18144}e^2
        +\frac{4041617}{2268}e^4
        +\frac{2242487}{1728}e^6
        -\frac{4337569}{14336}e^8
        -\frac{16318249}{258048}e^{10}
        +\left(1-e^2\right)^{3/2} \bigg(\frac{4429}{56}
        +\frac{43063}{112}e^2
        +\frac{31525}{64}e^4
        +\frac{53051}{896}e^6\bigg)\bigg], 
    \end{autobreak} \\
    \begin{autobreak}
        \MoveEqLeft
        \mathcal{A}_{7/2}^{S3} = q^3\bigg[
        -\frac{223}{12}
        -\frac{6941}{48}e^2
        -\frac{12581}{64}e^4
        -\frac{13269}{128}e^6
        -\frac{24343}{3072}e^8
        +6\left(1-e^2\right)^{3/2} \mathcal{A}_0 
        \bigg].
    \end{autobreak}
\end{align}
The remaining coefficient is a tail term with a non-terminating series in $e$, given below 
\begin{align}
    \begin{autobreak}
        \MoveEqLeft 
        \mathcal{A}_{7/2}^{S2} = q^2\pi\bigg( 
            \frac{65}{8}
            +\frac{2277}{32}e^2
            +\frac{103229}{768}e^4
            +\frac{1307875}{18432}e^6
            +\frac{2250275}{589824}e^8
            +\frac{12955751}{176947200}e^{10}
            -\frac{14777977}{849346560}e^{12}
            +\frac{912190597}{1109812838400}e^{14}
            -\frac{1342483733}{47352014438400}e^{16}
            -\frac{1419252436201}{27615694820474880}e^{18}
            -\frac{1808844112469}{187861869527040000}e^{20}
            + \cdots
        \bigg),
    \end{autobreak}
\end{align}
computed to $e^{20}$.

Advancing to the 4PN contributions, we find three non-zero coefficients with spin.  Two of these are exact 
expressions in $e$
\begin{align}
    \begin{autobreak}
        \MoveEqLeft
        \mathcal{A}_{4}^{S2} = q^2\bigg[
        \frac{3337877}{18144}
        +\frac{405437}{324}e^2
        +\frac{94722907}{72576}e^4
        +\frac{268741}{324}e^6
        +\frac{241118531}{258048}e^8
        +\frac{3276199}{32256}e^{10}
        -\left(1-e^2\right)^{3/2} \bigg(\frac{132817}{672}
        +\frac{1160015}{1344}e^2
        +\frac{193565}{256}e^4
        +\frac{866921}{10752}e^6\bigg)
    \bigg],
    \end{autobreak} \\
    \begin{autobreak}
        \MoveEqLeft
        \mathcal{A}_{4}^{S4} = q^4\bigg(
           \frac{41}{16}
        +\frac{79}{4}e^2
        +\frac{269}{12}e^4
        +\frac{2693}{192}e^6
        +\frac{7339}{6144}e^8
        -\frac{3}{2}\left(1-e^2\right)^{3/2} \mathcal{A}_0  
        \bigg).
    \end{autobreak}
\end{align}
The third is a non-terminating series in $e$
\begin{align}
    \begin{autobreak}
        \MoveEqLeft 
        \mathcal{A}_{4}^{S1} = q\pi\bigg(
            \frac{3389}{96}
            +\frac{57405}{224}e^2
            -\frac{15198859}{43008}e^4
            -\frac{59177891}{64512}e^6
            -\frac{25122496349}{49545216}e^8
            -\frac{3412842619}{137625600}e^{10}
            +\frac{5018155390867}{475634073600}e^{12}
            +\frac{365634707839}{59454259200}e^{14}
            +\frac{17805357662744539}{5114017559347200}e^{16}
            +\frac{6709307903485285333}{2899647956149862400}e^{18}
            +\frac{2521384303831489088179}{1546478909946593280000}e^{20}
            +\cdots
        \bigg).
    \end{autobreak}
\end{align}
Here we note that $\mathcal{A}_{4}^{S3} = 0$.  The same holds for $\mathcal{A}_{2}^{S1}$, which is absent from the 
2PN terms above, and extends to higher even PN order terms, with $\mathcal{A}_{i}^{S(i-1)} = 0$ for 
$i = \{2,4,6,8,10\}$, holding through 10PN. 

Proceeding to the 4.5PN contributions, we find four coefficients with spin.  We present the two exact expressions 
first,
\begin{align}
    \begin{autobreak}
        \MoveEqLeft 
        \mathcal{A}_{9/2L}^{S1} = -q\bigg(
           \frac{1369}{18}
        +\frac{79678}{63}e^2
        +\frac{3179761}{840}e^4
        +\frac{1719275}{504}e^6
        +\frac{13553357}{16128}e^8
        +\frac{107343}{4480}e^{10} 
        \bigg),
    \end{autobreak} \\
    \begin{autobreak}
        \MoveEqLeft 
        \mathcal{A}_{9/2}^{S3} = q^3\bigg[
           -\frac{8433}{28}
        -\frac{108707}{42}e^2
        -\frac{2508733}{672}e^4
        -\frac{1048203}{448}e^6
        -\frac{865831}{768}e^8
        -\frac{648787}{7168}e^{10}
        +\left(1-e^2\right)^{3/2} \bigg(\frac{37867}{168}
        +\frac{315961}{336}e^2
        +\frac{42951}{64}e^4
        +\frac{175823}{2688}e^6\bigg) 
        \bigg].
    \end{autobreak}
\end{align}
Beyond this point, constants such as $\pi^2$, $\log(2)$, $\log(3)$, and $\g_E$ begin to appear in some of the 
spin-dependent coefficients.  The remaining two 4.5PN coefficients have non-terminating expressions in $e$, which 
we present below 
\begin{align}
    \begin{autobreak}
        \MoveEqLeft 
        \mathcal{A}_{9/2}^{S1} = q\bigg[
           -\frac{343985009}{498960}
        +\frac{1369 \g_E }{9}
        -\frac{385 \pi ^2}{9}
        +\frac{95723 \log(2)}{315}
        +\bigg(-\frac{33607234229}{2328480} 
        +\frac{159356 \g_E }{63}
        -\frac{11695 \pi^2}{18}
        +\frac{33316 \log (2)}{105}
        +\frac{188811 \log (3)}{40}\bigg) e^2
        +\bigg(-\frac{361951335499}{8467200}
        +\frac{3179761 \g_E }{420}
        -\frac{21443 \pi^2}{12}
        +\frac{158977523 \log (2)}{1260}
        -\frac{130364883 \log (3)}{2240}\bigg) e^4
        +\bigg(-\frac{1005841269121}{22353408}
        +\frac{1719275 \g_E }{252}
        -\frac{115525 \pi^2}{72}
        -\frac{1619131561 \log (2)}{1260}
        +\frac{5626660383 \log(3)}{17920}
        +\frac{4982421875 \log (5)}{13824}\bigg)e^6
        +\bigg(-\frac{1449035920823}{59609088}
        +\frac{13553357 \g_E }{8064}
        -\frac{487355 \pi^2}{1152}
        +\frac{3086900646601 \log (2)}{362880}
        +\frac{14207516721 \log(3)}{16384}
        -\frac{39395017578125 \log (5)}{9289728}\bigg)e^8
        +\bigg(-\frac{2546262906169}{198696960}
        +\frac{107343 \g_E }{2240}
        -\frac{3321\pi^2}{256}
        -\frac{191003235106681 \log (2)}{4536000}
        -\frac{2280624020922231 \log(3)}{114688000}
        +\frac{1658023115234375 \log (5)}{74317824}
        +\frac{688360827060361 \log(7)}{88473600}\bigg) e^{10}
        + \cdots
        +\delta\mathcal{A}_{9/2}^{S1} e^{20}
        + \cdots\bigg],
    \end{autobreak} \\
    \begin{autobreak}
        \MoveEqLeft
        \mathcal{A}_{9/2}^{S2} = q^2\pi\bigg[
            \frac{78335}{672}
            +\frac{383469}{224}e^2
            +\frac{193007905}{43008}e^4
            +\frac{223706737}{64512}e^6
            +\frac{1871802943}{1835008}e^8
            +\frac{5252772017}{137625600}e^{10}
            -\frac{4401648976307}{475634073600}e^{12}
            -\frac{39979852472447}{6991820881920}e^{14}
            -\frac{38038153513232009}{11932707638476800}e^{16}
            -\frac{6150598557113758523}{2899647956149862400}e^{18}
            -\frac{66036272504274463063}{44185111712759808000}e^{20}
            + \cdots
        \bigg],
    \end{autobreak}
\end{align}
where $\gamma_E$ is Euler's constant.  Both expressions are computed up to $e^{20}$, but for 
$\mathcal{A}_{9/2}^{S1}$, we truncate the presentation at $e^{10}$ due to the complex structure of the expression.  
As a reminder that there are additional terms up to $e^{20}$ not presented, the $e^{20}$ coefficient is represented 
by $\delta\mathcal{A}_{9/2}^{S1}$.

At 5PN, there are five non-zero coefficients with spin.  We first present the two exact expressions 
\begin{align}
    \begin{autobreak}
        \MoveEqLeft
        \mathcal{A}_{5L}^{S2} = q^2\bigg(
               \frac{4601}{280}
        +\frac{67303}{336}e^2
        +\frac{127223}{224}e^4
        +\frac{1017463}{1344}e^6
        +\frac{833423}{3072}e^8
        +\frac{686619}{71680}e^{10}
        \bigg),
    \end{autobreak} \\
    \begin{autobreak}
        \MoveEqLeft 
        \mathcal{A}_{5}^{S4} = q^4\bigg[
           \frac{132019}{672}
        +\frac{1209181}{672}e^2
        +\frac{1147065}{448}e^4
        +\frac{7645439}{5376}e^6
        +\frac{7380797}{12288}e^8
        +\frac{230683}{5376}e^{10}
        -\left(1-e^2\right)^{3/2} \bigg(\frac{32843}{224}
        +\frac{268369}{448}e^2
        +\frac{95225}{256}e^4
        +\frac{357865}{10752}e^6\bigg) 
        \bigg].
    \end{autobreak}
\end{align}
The remaining contributions are represented by non-terminating series in $e$ 
\begin{align}
    \begin{autobreak}
        \MoveEqLeft
        \mathcal{A}_{5}^{S1} = q\pi\bigg(
        \frac{1049395}{3024}
        +\frac{85633087}{16128}e^2
        +\frac{3525482221}{193536}e^4
        +\frac{260806848791}{13934592}e^6
        +\frac{3911945775407}{668860416}e^8
        +\frac{200289353539571}{535088332800}e^{10}
        +\frac{773041703602277}{19263179980800}e^{12}
        +\frac{448388847636535481}{2517055517491200}e^{14}
        +\frac{4921289871098480419}{53697184373145600}e^{16}
        +\frac{1643310513177647423287}{29824950406112870400}e^{18}
        +\frac{775356107081832378686309}{20877465284279009280000}e^{20}
        + \cdots 
        \bigg),
    \end{autobreak} \\
    \begin{autobreak}
       \MoveEqLeft 
       \mathcal{A}_{5}^{S2} = q^2\bigg[
        -\frac{3560724457}{23284800}
        -\frac{4601 \g_E }{140}
        +\frac{43 \pi ^2}{4}
        -\frac{27499\log (2)}{420}
        +\bigg(\frac{55776212497}{13970880}
        -\frac{67303 \g_E}{168}
        +\frac{3145 \pi ^2}{24}
        +\frac{33277 \log (2)}{840}
        -\frac{234009 \log (3)}{280}\bigg) e^2
        +\bigg(\frac{27392397085}{7451136}
        -\frac{127223 \g_E}{112}
        +\frac{5945 \pi ^2}{16}
        -\frac{2599993 \log (2)}{144}
        +\frac{75584907 \log(3)}{8960}\bigg) e^4
        +\bigg(\frac{78720992519}{22353408}
        -\frac{1017463 \g_E}{672}
        +\frac{47545 \pi ^2}{96}
        +\frac{526540259 \log (2)}{4320}
        -\frac{565443747 \log(3)}{17920}
        -\frac{208984375 \log (5)}{6048}\bigg)e^6
        +\bigg(\frac{8438732671}{1548288}
        -\frac{833423 \g_E }{1536}
        +\frac{272615 \pi^2}{1536}
        -\frac{48448996199 \log (2)}{161280}
        -\frac{119997241101 \log(3)}{2293760}
        +\frac{668123046875 \log (5)}{4128768}\bigg)e^8
        +\bigg(\frac{3621875026757}{993484800}
        -\frac{686619 \g_E }{35840}
        +\frac{6417 \pi ^2}{1024}
        -\frac{27008760476791 \log (2)}{14515200}
        -\frac{5025957080607 \log(3)}{11468800}
        +\frac{18663349609375 \log (5)}{18579456}
        +\frac{34486555724663 \log(7)}{442368000}\bigg) e^{10}
        +\cdots 
        +\delta\mathcal{A}_{5}^{S2}e^{20}
       +\cdots\bigg],   
    \end{autobreak} \\
    \begin{autobreak}
        \MoveEqLeft 
        \mathcal{A}_{5}^{S3} = -q^3\pi\bigg(
           \frac{1375}{24}
        +\frac{25535}{32}e^2
        +\frac{535867}{256}e^4
        +\frac{99785323}{55296}e^6
        +\frac{334579751}{589824}e^8
        +\frac{212500939}{11796480}e^{10}
        -\frac{789812219}{157286400}e^{12}
        -\frac{683187720557}{221962567680}e^{14}
        -\frac{49434196850047}{28411208663040}e^{16}
        -\frac{68511439423521523}{59176488901017600}e^{18}
        -\frac{22511844330269171611}{27615694820474880000}e^{20} 
        +\cdots\bigg).
    \end{autobreak}
\end{align}
In the same manner as $\mathcal{A}_{9/2}^{S1}$, for $\mathcal{A}_{5}^{S2}$ we truncate the presentation to 
$e^{10}$ and use $\delta\mathcal{A}_{5}^{S2}$ as a reminder that the expression is computed up to $e^{20}$.

Finally, we present one additional term to illustrate new spin-dependent behavior.  
At 6.5PN order, there is a term that would normally be considered linear in spin, yet has added $q$ dependence.  
We find
\begin{align}
    \begin{autobreak}
        \MoveEqLeft 
        \mathcal{A}_{13/2}^{S1} = q\bigg[
            \bigg(\frac{128}{15} \Psi^{(0,2)}(q)
            +\frac{256 \log (\kappa )}{15}
            +\frac{1352455 \pi ^2}{2268}
            -\frac{2695926721 \gamma_E}{1746360}
            +\frac{1258752377510003}{157329572400}
            -\frac{4208517 \log (3)}{1760}
            +\frac{257721407 \log (2)}{1746360}\bigg)
            +\bigg(\frac{1912}{5} \Psi^{(0,2)}(q)
            +\frac{3824 \log (\kappa )}{5}
            +\frac{171585709 \pi ^2}{13608}
            -\frac{295616512943 \gamma_E}{15717240}
            +\frac{1299608253061297987}{11327729212800}
            +\frac{1011083984375 \log (5)}{16765056}
            +\frac{16480874061 \log (3)}{172480}
            -\frac{5492857134383\log (2)}{15717240}\bigg)e^2
            +\bigg(\frac{13653}{5} \Psi^{(0,2)}(q)
            +\frac{27306 \log (\kappa )}{5}
            +\frac{3107992903 \pi ^2}{54432}
            -\frac{2380543801873\gamma_E }{125737920}
            +\frac{4383860628954161813}{5663864606400}
            -\frac{791429931640625 \log (5)}{201180672}
            -\frac{22611016107 \log(3)}{344960}
            +\frac{1142126890291439 \log (2)}{125737920}\bigg)e^4
            +\bigg(\frac{25844}{5} \Psi^{(0,2)}(q)
            +\frac{51688 \log (\kappa)}{5}
            +\frac{11785757413 \pi ^2}{108864}
            -\frac{2770193869633 \gamma_E }{62868960}
            +\frac{139824234734862179119}{90621833702400}
            +\frac{50512933321530239\log (7)}{2956124160}
            +\frac{1127053486281546875 \log (5)}{14485008384}
            -\frac{283024127334177 \log (3)}{5519360}
            -\frac{9268649966540609 \log(2)}{62868960}\bigg)e^6
            +\bigg(\frac{22911}{8} \Psi^{(0,2)}(q)
            +\frac{22911 \log (\kappa )}{4}
            +\frac{80021582891 \pi ^2}{870912}
            -\frac{322430012980721 \gamma_E }{2011806720}
            +\frac{2245411211937451000123}{1449949339238400}
            -\frac{119808898694384316977 \log (7)}{189191946240}
            -\frac{1506892316700801921875 \log (5)}{1854081073152}
            +\frac{1408848061308958467 \log (3)}{1412956160}
            +\frac{37733880973418423687 \log (2)}{18106260480}\bigg)e^8
            +\bigg(\frac{6363}{16} \Psi^{(0,2)}(q)
            +\frac{6363 \log (\kappa )}{8}
            +\frac{50272903417 \pi ^2}{1741824}
            -\frac{199827883148627 \gamma_E}{2011806720}
            +\frac{300701407365010200337}{362487334809600}
            +\frac{93204685782449667160133 \log (7)}{9459597312000}
            +\frac{40355476551754551109375 \log(5)}{7416324292608}
            -\frac{1103943111627083930253 \log (3)}{141295616000}
            -\frac{12634935588668960489203 \log(2)}{452656512000}\bigg)e^{10}
        +\cdots 
        +\delta\mathcal{A}_{13/2}^{S1}e^{20} 
        +\cdots\bigg].
    \end{autobreak}
\end{align}
At this PN order and in this contribution, we begin to encounter non-polynomial spin contributions to the flux, 
coming in the form of $\kappa$, $\log\kappa$, and the following functions 
\begin{align}
    \Psi^{(r,\sigma)}(q) &= \psi^{(r)}\left(1+\frac{i\sigma q}{\kappa}\right) 
+ \psi^{(r)}\left(1-\frac{i\sigma q}{\kappa}\right), \\
    i\bar{\Psi}^{(r,\sigma)}(q) &= \psi^{(r)}\left(1+\frac{i\sigma q}{\kappa}\right) 
- \psi^{(r)}\left(1-\frac{i\sigma q}{\kappa}\right) , \\
  \tilde{\Psi}^{(r,\sigma)}(q) &= \Psi^{(r,\sigma)}(q) - 2\psi^{(r)}(1) .
\end{align}
Here $\psi^{(r)}(z)$ is the $r$-th order polygamma function and $\sigma$ is an integer.  For order $r=0$, 
$\tilde{\Psi}^{(0,\sigma)}(q) = H(i\sigma q/\kappa) + H(-i\sigma q/\kappa)$, where $H(z) = \psi(1+z) + \gamma_E$ is 
the (analytically continued) harmonic number function, so that 
$\tilde{\Psi}^{(0,\sigma)} = 2\zeta(3)\sigma^2 q^2 + \mathcal{O}(q^4)$ (generally,
$\tilde{\Psi}^{(r,\sigma)} = -\psi^{(r+2)}(1)\,\sigma^2 q^2 + \mathcal{O}(q^4)$),
$\bar{\Psi}^{(r,\sigma)} = \mathcal{O}(q)$, $\log\kappa \simeq \kappa -1 \simeq -q^2/2 + \mathcal{O}(q^4)$; all of 
these vanish as $a \to 0$.  This small $a$ behavior is important for the non-polynomial dependence that shows up in 
the $\mathcal{A}_{iLj}^{S0}$ coefficients at 8, 9, 9.5, and 10PN orders (but not at 8.5PN), guaranteeing that these 
terms, which enter only through $\tilde\Psi^{(r,\sigma)}$, $\bar\Psi^{(r,\sigma)}$, $\log\kappa$ and $\kappa-1$, 
make no contribution to the Schwarzschild limit.  The expressions are too lengthy to display here but can be 
extracted from the online dataset \cite{BHPTK18,UNCGrav22,PostNewtonianSelfForce}.

\subsection{Leading \texorpdfstring{$q^n$}{q\textasciicircum{}n} terms in the energy flux}

For the flux contributions with spin, we notice a pattern in the leading $q^n$ contributions.  Here we present the 
leading $q^n$ energy flux contributions up to 8PN 
\begin{align}
    \begin{autobreak}
        \MoveEqLeft
        \mathcal{A}_{3/2}^{S1} = -q\bigg(
            \frac{73}{12}
            +\frac{823}{24}e^2
            +\frac{949}{32}e^4
            +\frac{491}{192}e^6
        \bigg),
    \end{autobreak} \\
    \begin{autobreak} 
        \MoveEqLeft
        \mathcal{A}_{2}^{S2} = q^2\bigg(
            \frac{33}{16}
            +\frac{359}{32}e^2
            +\frac{1465}{128}e^4
            +\frac{883}{768}e^6
        \bigg),
    \end{autobreak} \\
    \begin{autobreak}
        \MoveEqLeft
        \mathcal{A}_{7/2}^{S3} = -q^3\bigg[
            \frac{223}{12}
            +\frac{6941}{48}e^2
            +\frac{12581}{64}e^4
            +\frac{13269}{128}e^6
            +\frac{24343}{3072}e^8
            -6\left(1-e^2\right)^{3/2}\mathcal{A}_{0}
        \bigg],
    \end{autobreak} \\
    \begin{autobreak}
        \MoveEqLeft 
        \mathcal{A}_{4}^{S4} = q^4\bigg[
            \frac{41}{16}
            +\frac{79}{4}e^2
            +\frac{269}{12}e^4
            +\frac{2693}{192}e^6
            +\frac{7339}{6144}e^8
            -\frac{3}{2}\left(1-e^2\right)^{3/2}\mathcal{A}_{0}
        \bigg],
    \end{autobreak} \\
    \begin{autobreak}
        \MoveEqLeft 
        \mathcal{A}_{11/2}^{S5} = -q^5\bigg[
            58
            +\frac{2115}{4}e^2
            +\frac{15091}{24}e^4
            +\frac{47723}{192}e^6
            +\frac{91889}{768}e^8
            +\frac{26603}{3072}e^{10}
            -\left(1-e^2\right)^{3/2}\bigg(\frac{103}{2}
            +207 e^2
            +\frac{3699}{32}e^4
            +\frac{613}{64}e^6\bigg)
        \bigg],
    \end{autobreak} \\
    \begin{autobreak}
        \MoveEqLeft 
        \mathcal{A}_{6}^{S6} = q^6\bigg[
            \frac{227}{32}
            +\frac{6127}{96}e^2
            +\frac{6091}{96}e^4
            +\frac{8105}{768}e^6
            +\frac{26175}{4096}e^8
            +\frac{493}{1024}e^{10}
            -\left(1-e^2\right)^{3/2}\bigg(\frac{227}{32}
            +\frac{5279}{192}e^2
            +\frac{10865}{768}e^4
            +\frac{587}{512}e^6\bigg)
        \bigg], 
    \end{autobreak} \\
    \begin{autobreak}
        \MoveEqLeft 
        \mathcal{A}_{15/2}^{S7} = -q^7\bigg(
            \frac{18905}{12}e^2
            +\frac{1431835}{384}e^4
            +\frac{992857}{768}e^6
            -\frac{322207}{6144}e^8
            -\frac{337379}{4096}e^{10}
            -\frac{366195}{8192}e^{12}
            -\frac{1099751}{49152}e^{14}
            -\frac{427613}{32768}e^{16}
            -\frac{26672221}{3145728}e^{18}
            -\frac{74238823}{12582912}e^{20}
            +\cdots
        \bigg),
    \end{autobreak} \\ 
    \begin{autobreak}
        \MoveEqLeft 
        \mathcal{A}_{8}^{S8} = q^8\bigg(
            \frac{2349}{16}e^2
            +\frac{82161}{256}e^4
            +\frac{102697}{1536}e^6
            -\frac{9299}{512}e^8
            -\frac{18307}{2048}e^{10}
            -\frac{330719}{98304}e^{12}
            -\frac{108893}{65536}e^{14}
            -\frac{1038547}{1048576}e^{16}
            -\frac{1025327}{1572864}e^{18}
            -\frac{959601}{2097152}e^{20}
            +\cdots\bigg).
    \end{autobreak}
\end{align}
For completeness, we give a similar presentation for the leading $q^n$ contributions for the angular momentum flux, 
relegated to Appendix~\ref{app:leadq}.  From the sequence, the odd-power $q^n$ terms appear at $(n+1/2)$PN orders, 
while the even-power $q^n$ terms appear at $n$PN orders.  This pattern was seen previously in inclined-spherical 
orbit fluxes \cite{CastETC25,CastETC25b}, but with inclination parameter dependence instead of eccentricity.

The apparent pattern in the leading $q^n$ terms leads naturally to the question of the predictability of 
higher-order $q^n$ terms.  For the non-spinning flux, the structure of the leading logarithmic ($\log p$) 
contributions to all PN orders is known from low-order multipole moment results in full PN theory 
\cite{MunnEvan19a,MunnEvan20a}.  Whether a similar infinite sequence of functions for the leading $q^n$ terms can 
be extracted from full PN theory is an open question.

\subsection{Numerical comparison}
\label{sec:numComp}

As a post-Newtonian expansion, our analytical expressions for the gravitational fluxes are valid for orbits with large
separations. However, we include an additional expansion in eccentricity, which further limits the validity of the flux
expressions toward lower eccentricity orbits. Mapping out the domain of validity of our gravitational flux expressions
is important, and to do so, it is useful to compare with numerical gravitational flux data. 

\begin{figure*}[htb!]
    \includegraphics[width=\textwidth]{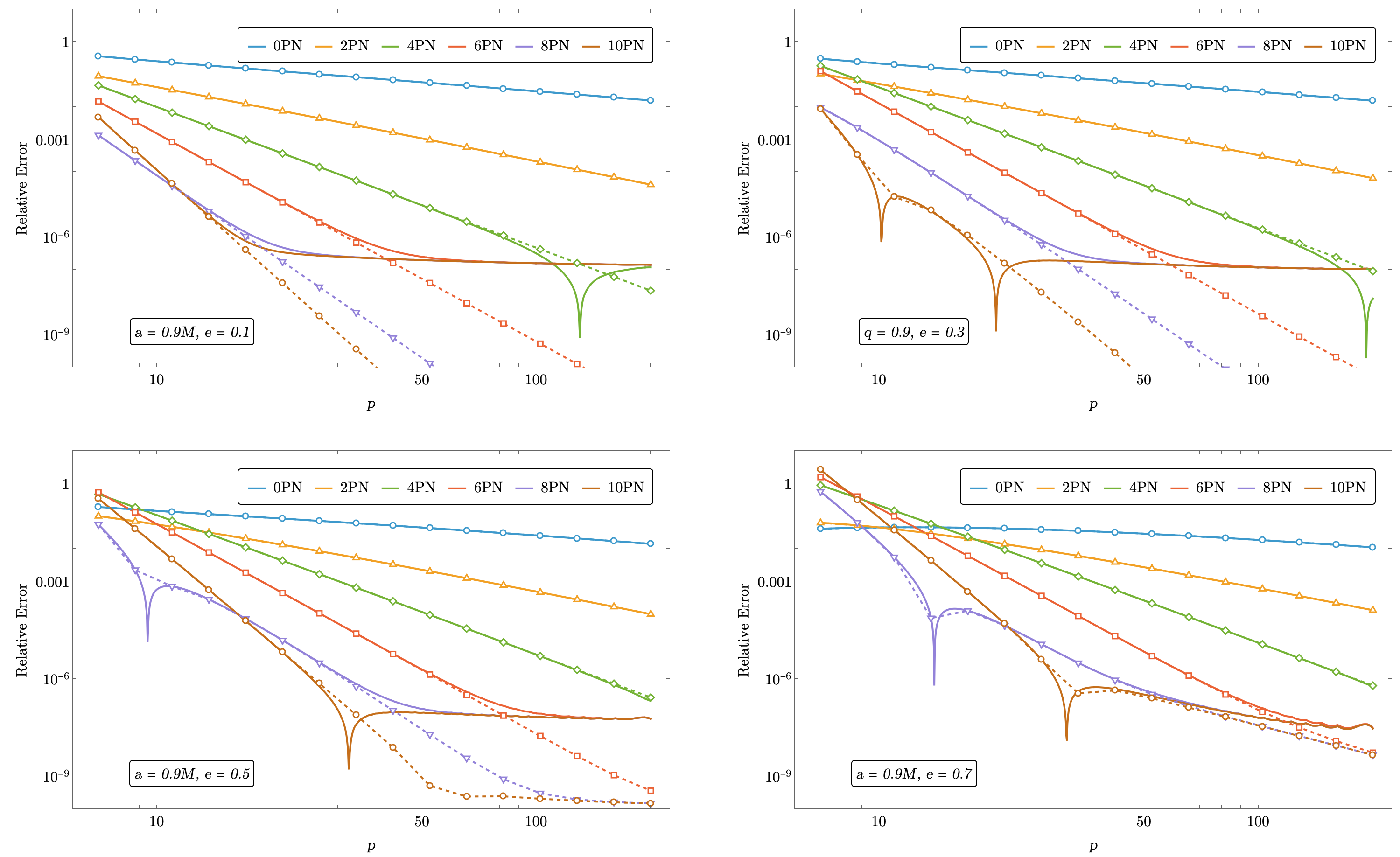}
    \caption{Relative error as a function of $p$ for the infinity-side gravitational energy flux 
$\langle\dot{E}\rangle_{\infty}$, presented as log-log plots, comparing the numerically evaluated expressions 
against numerical data, for orbital parameters $a = 0.9M$ and $e = \{0.1,0.3,0.5,0.7\}$.  The solid lines represent 
a comparison with interpolated flux data used in FEW~\cite{ChapBirdETC25}, while the dashed lines represent a 
comparison with higher-accuracy flux data.  The 
relative error for the higher PN order expressions eventually levels off as $p \to \infty$, indicating the dominance 
of the interpolation error of the numerical fluxes.  The region where the interpolation error dominates is pushed 
to larger $p$ as $e$ increases.  The sudden downward spikes occur in the sampled points that are near zero-crossings, 
$p_{C}$, which are locations where the PN and numerical energy fluxes fortuitously agree.  In the close separation, 
small $p$ regime, the 8PN expression is more accurate than the 10PN expression: for $p\lesssim11$ at $e=0.1$, where 
the $e^{20}$ truncation is negligible, consistent with an asymptotic PN series, and over a wider range 
($p\lesssim14$ and $p\lesssim21$) at $e=0.5$ and $0.7$.} 
\label{fig:numcompPN}
\end{figure*}

We compare our numerically evaluated PN fluxes with two sets of infinity-side fluxes: the interpolated fluxes used
by the FEW framework \cite{ChapBirdETC25}, and a discrete set of flux data computed with \texttt{pybhpt}, an
independent Teukolsky code \cite{pybhpt} (the FEW flux grid was itself largely computed with \texttt{pybhpt}
\cite{ChapBirdETC25}).  The FEW grid fluxes were computed to a requested relative precision of $10^{-8}$
\cite{ChapBirdETC25}.  After interpolation, we find that our comparisons level off at a relative error of about 
$10^{-7}$ in the large-separation region, where the PN error would otherwise continue its power-law falloff in $p$.  
Self-convergence of the PN expansion, examined by truncating at multiple PN orders, together with the comparison
against the discrete data below, indicates that for $e\lesssim0.3$ our 10PN series is accurate to better than
$10^{-10}$ beyond $p \simeq 50$.  At larger $e$, the $O(e^{22})$ truncation, which this test does not probe, limits
the accuracy instead.  The discrete set calculated by one of us (ZN) is more 
accurate: at $e=0.1$, our 10PN series agrees with it to $\sim10^{-13}$ at $p=200$.  This set 
therefore serves to demonstrate the performance of the PN fluxes in the region
where the comparison against the interpolated fluxes levels off.

In Fig.~\ref{fig:numcompPN}, we present a comparison for the infinity-side energy flux $\langle\dot{E}\rangle_\infty$ 
for $a = 0.9M$ and $e = \{0.1,0.3,0.5,0.7\}$.  The solid lines are the relative error values with the interpolated 
fluxes, while the hollow dots connected by dashed lines are the relative error values with the higher-accuracy flux 
data.  As seen in all four values for the eccentricity, the relative error falls off with increasing separation, with 
increasingly rapid falloff as more PN terms are included.  The interpolation error of the FEW 
data is reached earlier with more terms included; the discrete flux data set demonstrates the falloff at large 
separation more effectively. 

\begin{figure*}[htb!]
    \includegraphics[width=\textwidth]{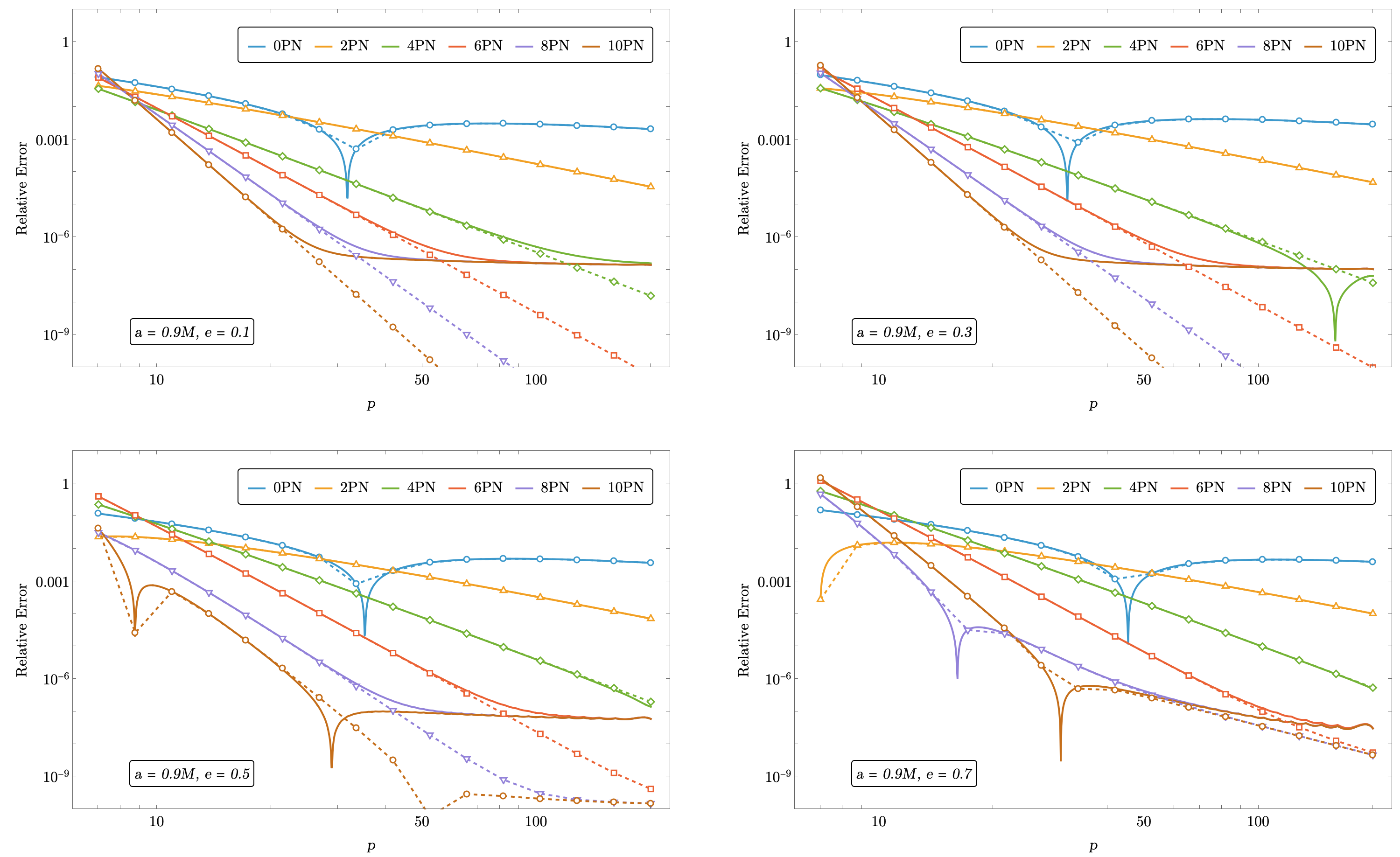} 
\caption{Relative error as a function of $p$ for the infinity-side average energy loss per radial period 
$\langle\Delta E\rangle_\infty = T_r\langle\dot E\rangle_\infty$, presented as log-log plots, once again 
comparing the numerically evaluated expressions against numerical flux data, for orbital parameters $a = 0.9M$ and 
$e = \{0.1,0.3,0.5,0.7\}$.  As with Fig.~\ref{fig:numcompPN}, the solid lines represent a comparison against the 
interpolated flux data used in FEW~\cite{ChapBirdETC25}, while the dashed lines represent a comparison against 
higher-accuracy flux data.  The higher-order behavior is similar to Fig.~\ref{fig:numcompPN}; 
for $e \lesssim 0.5$ the 10PN expressions remain more accurate than the 8PN expressions in close 
separation, a difference that arises from the PN truncation of $T_r$ (see text).}
    \label{fig:numcompPN_E_Loss}
\end{figure*}

A few other features stand out in Fig.~\ref{fig:numcompPN}.  The relative error curves are pushed outward as the 
eccentricity increases, decreasing the accuracy of the PN expansion at close separation.  In that region, 
the 8PN series is more accurate than the 10PN series for $p\lesssim11$ at $e=0.1$, where the $e^{20}$ truncation 
is negligible, revealing the asymptotic series nature of the PN expansion, 
and over a wider range ($p\lesssim14$ and $p\lesssim21$) at $e=0.5$ and $0.7$, where the eccentricity truncation 
also contributes.  We also note a few sharp downward spikes in the relative error in Fig.~\ref{fig:numcompPN}.  
These are the locations of zero-crossings, $p = p_{C}$, in the energy flux difference, where the fluxes fortuitously 
agree,
\begin{align}
  \left\langle\frac{dE}{dt}\right\rangle_\infty^\text{PN}
= \left\langle\frac{dE}{dt}\right\rangle_\infty^\text{Num} .
\end{align}
With sampled data, we are unlikely to hit a zero-crossing precisely, but near one the relative error in a 
log-log plot will have a sharp downward spike. 

The angular-momentum flux behaves similarly.  At $a = 0.9M$ its fractional residuals relative to the Teukolsky-code 
values range from about $3 \times 10^{-5}$ to $7 \times 10^{-3}$ at $p \simeq 10.9$, and from about 
$3 \times 10^{-12}$ to $5 \times 10^{-5}$ at $p \simeq 52.3$, for $e = 0.1$--$0.9$ (see the tutorial notebook in the 
online repositories \cite{UNCGrav22,BHPTK18}).

In Fig.~\ref{fig:numcompPN_E_Loss}, a similar comparison is made for
$\langle\Delta E\rangle_{\infty} = T_r \langle\dot{E}\rangle_{\infty}$, the energy radiated to infinity per radial
period.  Here $\langle\Delta E\rangle^{\rm PN}_\infty$ is the product of the PN flux and $T_r$ expanded to the same PN
order (exact in $e$), re-truncated at that order, while $\langle\Delta E\rangle^{\rm Num}_\infty$ uses the exact
$T_r$.  Were the exact $T_r$ used in both, the relative errors would coincide with those of
Fig.~\ref{fig:numcompPN}, so the differences between the two figures come from the PN truncation of $T_r$.  The
higher-order curves are similar; the 0PN curve differs, with a zero crossing, because the 1PN errors of the
Newtonian $T_r$ and of the leading-order flux partly cancel.  For $e \leq 0.5$, the 10PN series is more accurate
than the 8PN series in the region around $p\sim 10$--$20$, in contrast with the energy flux results, a difference
that likewise arises from the truncation of $T_r$.

The dependence of the relative error of the high PN order eccentric-equatorial flux expressions on the eccentricity
stands in contrast with the insensitivity of the analogous high PN order inclined-spherical flux expressions to
inclination \cite{CastETC25,SagoFujiNaka25}.  Two effects contribute to this sensitivity.  First, there is the 
truncation of the small-$e$ expansion, which introduces an $O(e^{22})$ error that competes with the PN truncation 
error and dominates that error at large $p$ for $e\gtrsim0.7$.  Second, although using $1/p$ as the PN compactness 
parameter keeps the pericenter distance
$r_{\rm min} = pM/(1+e)$ bounded below by $pM/2$, the local PN parameter at pericenter, $(1+e)/p$, still grows
by nearly a factor of two as $e\to1$, which slows the convergence of the PN series at small $p$.

\begin{figure*}[htb!]
    \includegraphics[width=\textwidth]{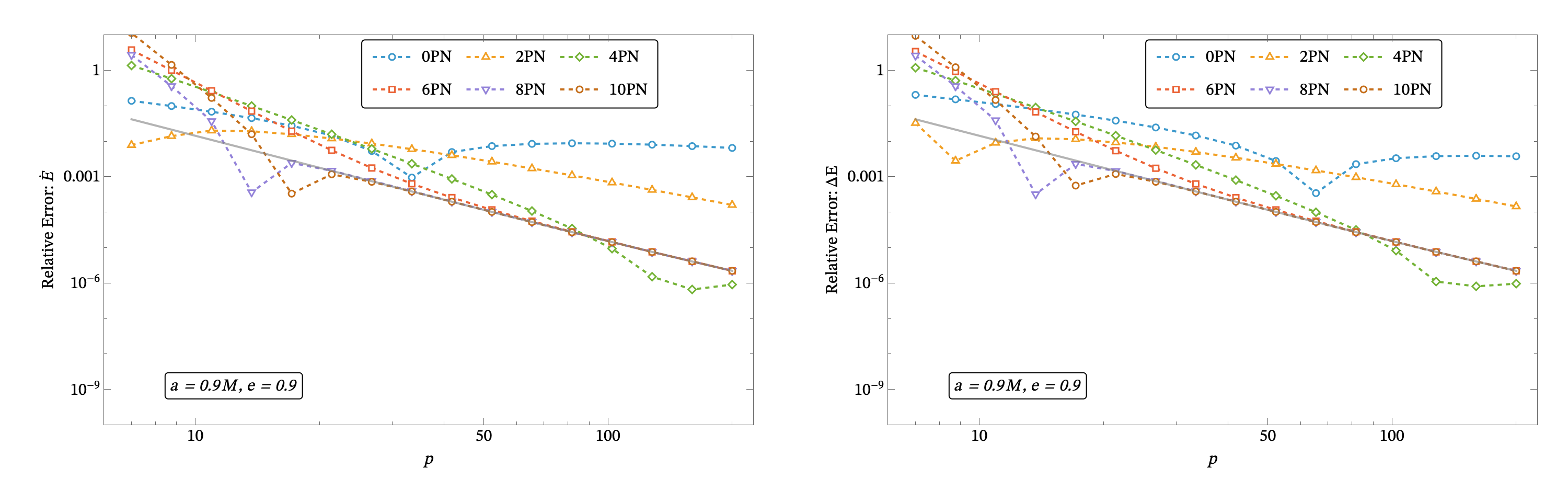}
    \caption{Relative error as a function of $p$ for the infinity-side $\dot{E}$ (left panel) and $\Delta E$ 
(right panel) presented as log-log plots, comparing the numerically evaluated expressions against numerical Teukolsky 
data from \texttt{pybhpt} \cite{pybhpt} for a highly 
eccentric orbit with parameters $a = 0.9M$ and $e = 0.9$.  The comparisons indicate that the $\geq 6$PN expressions 
converge to the same asymptotes for large $p$.  The gray semi-transparent lines represent fits for the asymptotes, 
with the dominant contribution over the range shown coming from the 3PN error.  In addition, for $4$PN, the relative 
error appears to undergo a zero-crossing at large $p$; we suspect that it will eventually converge with the 
relative errors of the $\geq 6$PN expressions.}
    \label{fig:numcompEC}
\end{figure*}

The competing error terms in PN and in $e$ reduce the utility of the high-order PN expansion as $e \to 1$.  
As a demonstration, in Fig.~\ref{fig:numcompEC}, we present a comparison of the energy flux and energy loss at 
infinity for $a = 0.9M$ and $e = 0.9$.  We still find power-law scaling for the residuals, with the $\geq 6$PN 
expressions converging to a single asymptote with power-law behavior; the $4$PN relative error, while seemingly 
of higher accuracy in the far-field region, appears to exhibit a zero-crossing, and we suspect that it will 
approach the same asymptote. We fitted this asymptote to the following error model
\begin{align}
    f_\text{err}(p) = \alpha_1 p^{-3} + \alpha_2 p^{-5/2} + \alpha_3 p^{-3/2},
\end{align}
based on where the first non-terminating terms appear (at 3PN, 2.5PN, and 1.5PN, respectively), represented by 
semi-transparent lines in Fig.~\ref{fig:numcompEC}.  Here we find $|\alpha_1| \gg |\alpha_2| \gg |\alpha_3|$, 
so that over the range shown the dominant error comes from 3PN, the first order at which a spin-dependent 
contribution has no known closed form and is known only through $e^{20}$.  The $p^{-3/2}$ term would dominate 
only beyond $p\sim400$.

\section{Conclusion}
\label{sec:conclusion}

In this work we presented analytic expressions for the gravitational fluxes radiated to infinity from 
EMRIs with eccentric-equatorial orbits about a Kerr black hole primary as expansions up to 10PN order and order 
$e^{20}$ in eccentricity.  The resulting expressions are exact in spin $a$ for $|a|<M$.  Individual flux 
coefficients are presented and discussed up to 5PN along with select higher-order terms.  We specifically 
highlighted the spin-dependent flux coefficients and their eccentricity dependence, as well as the leading 
$q^n$ terms present in the eccentric-equatorial flux.  

Many of the spin-dependent flux components are shown to resum into closed-form expressions in $e$.  These include 
the leading-spin terms of the energy flux through 6PN and of the angular momentum flux through 10PN, with the single
exception of $\mathcal{C}_{19/2}^{S9}$, which likely does have a closed form but cannot be established with our 
$e^{20}$ expansions.  We also find that certain terms vanish identically, with $\mathcal{A}_{i}^{S(i-1)} = 0$ 
for $i = 2,4,\dots,10$.  A 6.5PN term displays the onset of non-polynomial spin dependence that is then present in 
many higher-order PN terms.  The complete dataset can be obtained from online repositories 
\cite{BHPTK18,UNCGrav22,PostNewtonianSelfForce}.  

For the comparison of the PN-expanded fluxes with numerical flux data, the 10PN relative error falls below $10^{-7}$
for $p\gtrsim20$ ($e=0.1$), $p\gtrsim35$ ($e=0.5$) and $p\gtrsim75$ ($e=0.7$).  At $e=0.7$ the 8PN and 10PN errors
coincide for $p\gtrsim50$, so there the $O(e^{22})$ truncation, not the PN order, sets the accuracy.  This 
truncation error also makes the relative error curves sensitive to the value of $e$.  High values of $e$ demonstrate 
a slower falloff rate, with worse behavior at small $p$, toward the separatrix.  This $O(e^{22})$ error eventually 
causes a breakdown in the expected behavior of the relative error curves as $e \to 1$, as the residuals begin to 
converge to a single power-law asymptote.

A next step will be publishing the comparable expansions for the energy and angular momentum radiated into the 
horizon of the primary black hole.  For future work, we are interested in using the combined PN flux expressions 
and individual mode amplitude expansions to generate adiabatic waveforms from eccentric-equatorial EMRIs.  An 
effort will be made to improve the performance of the PN-expanded fluxes in the $e \gtrsim 0.7$ regime.  EMRIs may 
form with high initial eccentricity \cite{HopmAlex05}, with a recent study suggesting 
some EMRIs may retain a high eccentricity even at the point of plunge \cite{MancETC25}.  While a larger expansion 
in $e$ is always possible, a recent calculation for Schwarzschild eccentric fluxes \cite{KhalKavaTele26} has 
demonstrated the feasibility of a method without an additional expansion in $e$.     

\section*{Acknowledgments}
We thank Scott Hughes, Soichiro Isoyama, Barry Wardell, Jakob Neef, Chris Kavanagh, and Adrian Ottewill for useful 
discussions.  This research was sponsored by NSF Grant Nos.~PHY-2110335 and PHY-2409604 to the University of North 
Carolina at Chapel Hill and the Hamilton Award, University of North Carolina at Chapel Hill.  JCC acknowledges 
support from the European Union's Horizon Europe research and innovation program under the Marie 
Sk{\csname OT1\string\l\endcsname}odowska-Curie grant
agreement No.~101282032 during manuscript revision and final preparation of this paper.  ZN acknowledges the 
support of the ERC Consolidator/UKRI Frontier Research Grant GWModels (selected by the ERC and funded by UKRI [grant 
number EP/Y008251/1]). 

\textit{AI usage statement:} Claude Opus 5.5 and Claude Code (Opus 5.5) by Anthropic were used through IrisMetis, 
a research-focused AI harness developed by one of us (CRE).  This tool was used (i) to check equations in the 
formalism against cited sources, including performing independent symbolic and high-precision numerical 
verifications, (ii) to check our presented results against our original \textsc{Mathematica} data, (iii) to check 
references, and (iv) to make a critical review of the manuscript.  All scientific analyses, analytic and numerical 
results, and conclusions were produced by the authors, who are responsible for the content of this work.

\appendix

\section{Angular momentum flux contributions}\label{app:AngMomFlux}

For the angular momentum flux contributions, in the same manner as the energy flux contributions, we note first 
the leading-order contribution from the results of Peters \cite{Pete64}
\begin{align}
    \mathcal{C}_{0} = 1+\frac{7}{8}e^2,
\end{align}
which reappears in some of the higher-order contributions. As with the energy flux, the lowest-order 
contribution with spin appears at 1.5PN
\begin{align}
    \mathcal{C}_{3/2}^{S1} = -q \left(\frac{61}{12}+\frac{119}{8}e^2+\frac{183}{32}e^4\right).
\end{align}
The next-to-lowest-order contribution with spin appears immediately at 2PN 
\begin{align}
    \mathcal{C}_{2}^{S2} = q^2\left(\frac{33}{16}+\frac{95}{16}e^2+\frac{311}{128}e^4\right),
\end{align}
and then the one after appears at 2.5PN 
\begin{align}
    \begin{autobreak} 
    \MoveEqLeft
    \mathcal{C}_{5/2}^{S1} = 
    -q\bigg(\frac{255}{56}
    +\frac{401}{672}e^2
    +\frac{827}{24}e^4
    +\frac{178805}{5376}e^6
    -12\left(1-e^2\right)^{3/2} \mathcal{C}_{0}\bigg).
    \end{autobreak}
\end{align}
Up to this point, the contributions with spin can be expressed as exact functions in $e$.  Starting with 3PN, 
contributions with spin begin to appear with non-terminating series expansions in $e$ 
\begin{align}
    \begin{autobreak}
        \MoveEqLeft 
        \mathcal{C}_{3}^{S1} = -q\pi\bigg(
            \frac{145}{6}
            +\frac{409}{3}e^2
            +\frac{22631}{192}e^4
            +\frac{69887}{6912}e^6
            -\frac{102053}{442368}e^8
            -\frac{156871}{11059200}e^{10}
            -\frac{498149}{353894400}e^{12}
            +\frac{1230743}{416179814400}e^{14}
            -\frac{73068577}{608811614208}e^{16}
            -\frac{33675269011}{404526779596800}e^{18}
            -\frac{893949397653031}{20711771115356160000}e^{20}
            +\cdots
        \bigg),
    \end{autobreak}
\end{align}
along with a contribution with an exact expression in $e$ 
\begin{align}
    \begin{autobreak}
        \MoveEqLeft 
        \mathcal{C}_{3}^{S2} = q^2\bigg(
            \frac{1415}{56}
            +\frac{1457}{14}e^2
            +\frac{129299}{1344}e^4
            +\frac{4093}{112}e^6
            -11\left(1-e^2\right)^{3/2}\mathcal{C}_{0}
        \bigg).
    \end{autobreak}
\end{align}

At 3.5PN there are three contributions with spin, two of which are exact expressions in $e$ 
\begin{align}
    \begin{autobreak}
        \MoveEqLeft 
        \mathcal{C}_{7/2}^{S1} = -q\bigg[
            \frac{1279}{108}
            -\frac{4125061}{18144}e^2
            -\frac{14788903}{24192}e^4
            -\frac{6251351}{48384}e^6
            +\frac{17416811}{129024}e^8
            -\left(1-e^2\right)^{3/2}\bigg(\frac{4009}{56}
            +\frac{24487}{112}e^2
            +\frac{61885}{448}e^4\bigg)
        \bigg],
    \end{autobreak} \\
    \begin{autobreak}
        \MoveEqLeft 
        \mathcal{C}_{7/2}^{S3} = -q^3\bigg(
            \frac{793}{48}
            +\frac{1165 e^2}{16}
            +\frac{7345 e^4}{128}
            +\frac{1031 e^6}{64}
            -6\left(1-e^2\right)^{3/2}\mathcal{C}_{0}
        \bigg),
    \end{autobreak}
\end{align}
and the remaining contribution is a non-terminating series expansion in $e$ 
\begin{align}
    \begin{autobreak}
        \MoveEqLeft 
        \mathcal{C}_{7/2}^{S2} = q^2\pi\bigg(
            \frac{65}{8}
            +\frac{705}{16}e^2
            +\frac{11531}{256}e^4
            +\frac{10525}{2304}e^6
            -\frac{4485}{65536}e^8
            -\frac{360349}{29491200}e^{10}
            -\frac{1414283}{1415577600}e^{12}
            +\frac{4301621}{92484403200}e^{14}
            -\frac{14268745469}{426168129945600}e^{16}
            -\frac{426405514313}{13807847410237440}e^{18}
            -\frac{153418518516841}{9205231606824960000}e^{20}
            +\cdots
        \bigg).
    \end{autobreak}
\end{align}

Proceeding to 4PN, there are three contributions with spin, two of which are exact expressions in $e$ 
\begin{align}
    \begin{autobreak}
        \MoveEqLeft 
        \mathcal{C}_{4}^{S2} = q^2\bigg[
            \frac{3322595}{18144}
            +\frac{24624379}{36288}e^2
            +\frac{6311405}{16128}e^4
            +\frac{29684975}{96768}e^6
            +\frac{28287899}{129024}e^8
            -\left(1-e^2\right)^{3/2} \bigg(\frac{124753}{672}
            +\frac{270055}{672}e^2
            +\frac{992711}{5376}e^4\bigg)
        \bigg],
    \end{autobreak} \\
    \begin{autobreak}
        \MoveEqLeft 
        \mathcal{C}_{4}^{S4} = q^4\bigg(
            \frac{41}{16}
            +\frac{91}{8}e^2
            +\frac{1029}{128}e^4
            +\frac{143}{64}e^6
            -\frac{3}{2}\left(1-e^2\right)^{3/2}\mathcal{C}_{0}
        \bigg),
    \end{autobreak}
\end{align}
and the remaining contribution is a non-terminating series expansion in $e$ 
\begin{align}
    \begin{autobreak}
        \MoveEqLeft 
        \mathcal{C}_{4}^{S1} = q\pi\bigg(
            \frac{3883}{168}
            -\frac{21013}{336}e^2
            -\frac{269503}{448}e^4
            -\frac{110189735}{193536}e^6
            -\frac{586759099}{12386304}e^8
            +\frac{2077277633}{206438400}e^{10}
            +\frac{3983964959}{1698693120}e^{12}
            +\frac{16789016789551}{11653034803200}e^{14}
            +\frac{55551336774499}{60881161420800}e^{16}
            +\frac{909344150388440029}{1449823978074931200}e^{18}
            +\frac{131223085626576791321}{289964795614986240000}e^{20}
            +\cdots
        \bigg).
    \end{autobreak}
\end{align}

Logarithmic contributions with spin begin to appear at 4.5PN.  At that order, there are four contributions with 
spin, two of which are exact expressions in $e$ 
\begin{align}
    \begin{autobreak}
        \MoveEqLeft 
        \mathcal{C}_{9/2L}^{S1} = -q\bigg(
            \frac{42779}{630}
            +\frac{147127}{210}e^2
            +\frac{282671}{240}e^4
            +\frac{1140829}{2520}e^6
            +\frac{17153}{896}e^8
        \bigg),
    \end{autobreak} \\
    \begin{autobreak}
        \MoveEqLeft 
        \mathcal{C}_{9/2}^{S3} = -q^3\bigg[
            \frac{45331}{168}
            +\frac{283911}{224}e^2
            +\frac{984433}{896}e^4
            +\frac{2985271}{5376}e^6
            +\frac{2702285}{14336}e^8
            -\left(1-e^2\right)^{3/2} \bigg(\frac{36019}{168}
            +\frac{5597}{14}e^2
            +\frac{66107}{448}e^4\bigg)
        \bigg], 
    \end{autobreak}
\end{align} 
and the remaining two contributions are non-terminating series expansions in $e$ 
\begin{align}
    \begin{autobreak}
        \MoveEqLeft 
        \mathcal{C}_{9/2}^{S1} = -q\bigg[
            \bigg(\frac{41514161741}{69854400}
            -\frac{42779 \gamma_E }{315}
            +\frac{337 \pi^2}{9}
            -\frac{85451 \log(2)}{315}\bigg)
            +\bigg(\frac{60982819811}{7761600}
            -\frac{147127 \gamma_E}{105}
            +\frac{719 \pi^2}{2}
            +\frac{151211 \log (2)}{315}
            -\frac{907119 \log (3)}{280}\bigg)e^2
            +\bigg(\frac{6231009899}{475200}
            -\frac{282671 \gamma_E}{120}
            +\frac{2331 \pi^2}{4}
            -\frac{174683083 \log (2)}{2520}
            +\frac{16094133 \log (3)}{448}\bigg)e^4
            +\bigg(\frac{1442919678239}{186278400}
            -\frac{1140829 \gamma_E}{1260}
            +\frac{33503 \pi^2}{144}
            +\frac{2375549881 \log (2)}{3780}
            -\frac{3049985097 \log (3)}{17920}
            -\frac{733984375 \log (5)}{4608}\bigg)e^6
            +\bigg(\frac{2343131727719}{596090880}
            -\frac{17153 \gamma_E }{448}
            +\frac{1345 \pi^2}{128}
            -\frac{637741071997 \log (2)}{181440}
            -\frac{135126305919 \log (3)}{573440}
            +\frac{2219027734375 \log (5)}{1327104}\bigg)e^8
            +\bigg(\frac{36163633301}{14192640}
            +\frac{8345969720873 \log (2)}{567000}
            +\frac{752213879312949 \log (3)}{114688000}
            -\frac{576846620703125 \log (5)}{74317824}
            -\frac{372715074520789 \log (7)}{147456000}\bigg)e^{10}
            +\cdots 
            +\delta\mathcal{C}_{9/2}^{S1}e^{20}
            +\cdots\bigg],
    \end{autobreak} \\
    \begin{autobreak}
        \MoveEqLeft 
        \mathcal{C}_{9/2}^{S2} = q^2\pi\bigg(
            \frac{59407}{672}
            +\frac{45707}{48}e^2
            +\frac{1631583}{1024}e^4
            +\frac{623286169}{774144}e^6
            +\frac{385232489}{7077888}e^8
            -\frac{1806826607}{206438400}e^{10}
            -\frac{106920586889}{47563407360}e^{12}
            -\frac{2472833670961}{1864485568512}e^{14}
            -\frac{5997303750007027}{7159624583086080}e^{16}
            -\frac{119186910180983777}{207117711153561600}e^{18}
            -\frac{192549139708655958877}{463943672983977984000}e^{20}
            +\cdots
        \bigg).
    \end{autobreak}
\end{align}
Here we use the same shorthand notation used for energy flux contributions that become exceedingly complex, 
showing only the terms up to $e^{10}$ and using $\delta\mathcal{C}_{iLj}^{Sk}$ as a marker that the expression 
is computed up to a specified $e^{n}$. 

Finally, we present the 5PN contributions.  There are five contributions with spin; two have exact expressions 
in $e$, and the remaining three are expressed as a non-terminating expansion in eccentricity.  The contributions 
with exact expressions are 
\begin{align}
    \begin{autobreak}
        \MoveEqLeft 
        \mathcal{C}_{5L}^{S2} = q^2\bigg(
            \frac{4601}{280}
            +\frac{59813}{420}e^2
            +\frac{900191}{3360}e^4
            +\frac{31779}{224}e^6
            +\frac{273171}{35840}e^8
        \bigg),
    \end{autobreak} \\
    \begin{autobreak}
        \MoveEqLeft 
        \mathcal{C}_{5}^{S4} = q^4\bigg[
            \frac{119531}{672}
            +\frac{58635}{64}e^2
            +\frac{3986189}{5376}e^4
            +\frac{3310505}{10752}e^6
            +\frac{1188273}{14336}e^8
            -\left(1-e^2\right)^{3/2}\bigg(\frac{31499}{224}
            +\frac{54105}{224}e^2
            +\frac{131893}{1792}e^4\bigg)
        \bigg].
    \end{autobreak}    
\end{align}
The remaining non-terminating contributions are then 
\begin{align}
    \begin{autobreak}
        \MoveEqLeft 
        \mathcal{C}_{5}^{S1} = q\pi\bigg(
            \frac{135955}{432}
            +\frac{19964015}{6048}e^2
            +\frac{473602823}{64512}e^4
            +\frac{3552960917}{1306368}e^6
            -\frac{1064470959997}{668860416}e^8
            -\frac{75347499679}{2477260800}e^{10}
            +\frac{882332713411741}{6421059993600}e^{12}
            +\frac{3910479578337551}{94389581905920}e^{14}
            +\frac{12652032226152736001}{483274659358310400}e^{16}
            +\frac{1409749478119137349}{82847084461424640}e^{18}
            +\frac{82431757815679195661707}{6959155094759669760000}e^{20}
            +\cdots
        \bigg),
    \end{autobreak} \\
    \begin{autobreak}
        \MoveEqLeft 
        \mathcal{C}_{5}^{S2} = q^2\bigg[
            \bigg(-\frac{6164329271}{69854400}
            -\frac{4601 \gamma_E }{140}
            +\frac{43 \pi ^2}{4}
            -\frac{27499 \log (2)}{420}\bigg)
            +\bigg(\frac{64309909487}{17463600}
            -\frac{59813 \gamma_E }{210}
            +\frac{559 \pi ^2}{6}
            +\frac{1391 \log (2)}{10}
            -\frac{78003 \log (3)}{112}\bigg)e^2
            +\bigg(\frac{115696071307}{31046400}
            -\frac{900191 \gamma_E }{1680}
            +\frac{8413 \pi ^2}{48}
            -\frac{21851647 \log (2)}{1680}
            +\frac{8502327 \log (3)}{1280}\bigg) e^4
            +\bigg(\frac{14246534311}{5080320}
            -\frac{31779 \gamma_E }{112}
            +\frac{1485 \pi ^2}{16}
            +\frac{1371209387 \log (2)}{15120}
            -\frac{875739681 \log (3)}{35840}
            -\frac{4555859375 \log (5)}{193536}\bigg) e^6
            +\bigg(\frac{5050644835543}{2980454400}
            -\frac{273171 \gamma_E }{17920}
            +\frac{2553 \pi ^2}{512}
            -\frac{31166491661 \log (2)}{96768}
            -\frac{16857930357 \log (3)}{458752}
            +\frac{2020586328125 \log (5)}{12386304}\bigg)e^8
            +\bigg(\frac{47315532431}{70963200}
            +\frac{6459775859 \log (2)}{18900}
            +\frac{2253054018591 \log (3)}{6553600}
            -\frac{3248076953125 \log (5)}{16515072}
            -\frac{135800258431999 \log (7)}{884736000}\bigg) e^{10}
            +\cdots 
            +\delta\mathcal{C}_{5}^{S2}e^{20}
            +\cdots
        \bigg],
    \end{autobreak} \\
    \begin{autobreak}
        \MoveEqLeft 
        \mathcal{C}_{5}^{S3} = -q^3\pi\bigg(
            \frac{295}{6}
            +\frac{12611}{24}e^2
            +\frac{161513}{192}e^4
            +\frac{2696219}{6912}e^6
            +\frac{10202275}{442368}e^8
            -\frac{180508039}{44236800}e^{10}
            -\frac{16624343}{13107200}e^{12}
            -\frac{149544663293}{208089907200}e^{14}
            -\frac{48623023252381}{106542032486400}e^{16}
            -\frac{32487587836067527}{103558855576780800}e^{18}
            -\frac{669481925807826283}{2958824445050880000}e^{20}
        +\cdots\bigg).
    \end{autobreak}
\end{align}
Once again we truncated the expression for $\mathcal{C}_{5}^{S2}$ to $e^{10}$ and used 
$\delta\mathcal{C}_{5}^{S2}$ as a signifier that additional terms are computed up to $e^{20}$. 

\section{Leading \texorpdfstring{$q^n$}{q\textasciicircum{}n} terms in the angular momentum flux}\label{app:leadq}

This appendix lists the leading-spin terms in the angular momentum flux through 10PN order.  All but one of 
these resum to closed-form expressions.  The single exception is the $\mathcal{C}_{19/2}^{S9}$ term,
whose computed series through $e^{20}$ has too few terms to determine whether it has a closed form, although the 
pattern of the neighboring terms suggests one exists.  It should be apparent from the number of terms shown in this 
section that it is easier to find closed-form expressions in the angular momentum fluxes than in the energy flux.
\begin{align}
    \begin{autobreak} 
        \MoveEqLeft
        \mathcal{C}_{3/2}^{S1} = -q\bigg(
            \frac{61}{12}
            +\frac{119}{8}e^2
            +\frac{183}{32}e^4\bigg),
    \end{autobreak} \\ 
    \begin{autobreak}
        \MoveEqLeft 
        \mathcal{C}_{2}^{S2} = q^2\bigg(
            \frac{33}{16}
            +\frac{95}{16}e^2
            +\frac{311}{128}e^4
        \bigg),
    \end{autobreak} \\
    \begin{autobreak}
        \MoveEqLeft
        \mathcal{C}_{7/2}^{S3} = -q^3\bigg(
            \frac{793}{48}
            +\frac{1165}{16}e^2
            +\frac{7345}{128}e^4
            +\frac{1031}{64}e^6
            -6\left(1-e^2\right)^{3/2}\mathcal{C}_{0}
        \bigg),
    \end{autobreak} \\
    \begin{autobreak}
        \MoveEqLeft 
        \mathcal{C}_{4}^{S4} = q^4\bigg(
            \frac{41}{16}
            +\frac{91}{8}e^2
            +\frac{1029}{128}e^4
            +\frac{143}{64}e^6
            -\frac{3}{2}\left(1-e^2\right)^{3/2}\mathcal{C}_{0}
        \bigg),
    \end{autobreak} \\
    \begin{autobreak}
        \MoveEqLeft 
        \mathcal{C}_{11/2}^{S5} = -q^5\bigg[
            \frac{887}{16}
            +\frac{3479}{12}e^2
            +\frac{71851}{384}e^4
            +\frac{5785}{96}e^6
            +\frac{15439}{1024}e^8
            -\left(1-e^2\right)^{3/2} \bigg(50
            +\frac{1299}{16}e^2
            +\frac{657}{32}e^4\bigg)\bigg],
    \end{autobreak} \\
    \begin{autobreak}
        \MoveEqLeft 
        \mathcal{C}_{6}^{S6} = q^6\bigg[
            \frac{227}{32}
            +\frac{2361}{64}e^2
            +\frac{4475}{256}e^4
            +\frac{1743}{512}e^6
            +\frac{87}{128}e^8
            -\left(1-e^2\right)^{3/2} \bigg(\frac{227}{32}
            +\frac{349}{32}e^2
            +\frac{597}{256}e^4\bigg)
        \bigg],
    \end{autobreak} \\
    \begin{autobreak}
        \MoveEqLeft 
        \mathcal{C}_{15/2}^{S7} = -q^7\bigg[
            \frac{22403}{96}
            +\frac{80399}{64}e^2
            +\frac{202743}{256}e^4
            +\frac{192223}{1536}e^6
            +\frac{7779}{2048}e^8
            +\frac{2407}{512}e^{10}
            -\left(1-e^2\right)^{3/2} \bigg(\frac{22403}{96}
            +\frac{14557}{32}e^2
            +\frac{39651}{256}e^4
            +\frac{1557}{64}e^6\bigg)
        \bigg],
    \end{autobreak} \\
    \begin{autobreak}
        \MoveEqLeft 
        \mathcal{C}_{8}^{S8} = q^8\bigg[
            \frac{735}{32}
            +\frac{3755}{32}e^2
            +\frac{14537}{256}e^4
            +\frac{375}{256}e^6
            -\frac{63}{32}e^8
            -\left(1-e^2\right)^{3/2} \bigg(\frac{735}{32}
            +\frac{2651}{64}e^2
            +\frac{2377}{256}e^4
            +\frac{261}{256}e^6\bigg)\bigg] ,
    \end{autobreak} \\
\everyafterautobreak{\rule{0pt}{4.5ex}}
    \begin{autobreak}
        \MoveEqLeft
        \mathcal{C}_{19/2}^{S9} = -q^9\bigg[
            \frac{69405}{16}e^2
            +\frac{1461117}{256}e^4
            +\frac{705875}{1536}e^6
            -\frac{1783217}{4096}e^8
            -\frac{304479}{2048}e^{10}
            -\frac{2693335}{49152}e^{12}
            -\frac{1882187}{65536}e^{14}
            -\frac{9331969}{524288}e^{16}
            -\frac{37778015}{3145728}e^{18}
            -\frac{4497015}{524288}e^{20}+\cdots\bigg],
    \end{autobreak} \\
    \begin{autobreak}
        \MoveEqLeft 
        \mathcal{C}_{10}^{S10} = q^{10}\bigg[
            \frac{9385}{128}
            +\frac{12231}{32}e^2
            +\frac{56481}{256}e^4
            +\frac{3417}{512}e^6
            -\frac{198827}{16384}e^8
            -\frac{783}{512}e^{10}
            -\left(1-e^2\right)^{3/2} \bigg(\frac{9385}{128}
            +\frac{41363}{256}e^2
            +\frac{29991}{1024}e^4
            +\frac{55}{2048}e^6\bigg)\bigg] .
    \end{autobreak}
\end{align}
\bibliography{references}
\end{document}